\documentclass[a4paper,10pt]{article}
\pdfoutput=1
\usepackage{jcappub}
\usepackage{mathtools}
\usepackage{ulem}
\usepackage{bm}
\usepackage{xcolor}
\usepackage{natbib}
\usepackage{orcidlink}
\usepackage [latin1]{inputenc}
\usepackage{graphicx}
\usepackage{subcaption}

\def\laq{~\raise 0.4ex\hbox{$<$}\kern -0.8em\lower 0.62ex\hbox{$\sim$}~}
\def\gaq{~\raise 0.4ex\hbox{$>$}\kern -0.7em\lower 0.62ex\hbox{$\sim$}~}

\def\beq{\begin{equation}}
\def\eeq{\end{equation}}
\def\bea{\begin{eqnarray}}
\def\eea{\end{eqnarray}}

\title{Freeze-out and spectral running of primordial gravitational waves in viscous cosmology}
\author[a]{Giuseppe Fanizza \orcidlink{0000-0001-5173-3800},}
\author[b,c]{Eliseo Pavone \orcidlink{0000-0002-3022-4545},}
\author[b,c] {Luigi Tedesco \orcidlink{0000-0001-6508-2658}}    
\affiliation[a]{Dipartimento di Ingegneria, Libera Universit\`a Mediterranea (LUM) ``Giuseppe Degennaro'', S.S. 100 km 18, 70010 Casamassima, Bari, Italy}
\affiliation[b]{Dipartimento di Fisica, Universit\`a di Bari, Via G. Amendola 173, 70126 Bari, Italy}
\affiliation[c]{Istituto Nazionale di Fisica Nucleare, Sezione di Bari, Italy}
\emailAdd{fanizza@lum.it}
\emailAdd{eliseo.pavone@ba.infn.it}
\emailAdd{luigi.tedesco@ba.infn.it}

\abstract{We investigate the impact of shear viscosity on the propagation of primordial gravitational waves (pGW) after inflation.
Without assuming a specific inflationary scenario we focus on the evolution of  pGWs  after they re-enter the horizon during a cosmological epoch characterized by the presence of  shear viscosity. We show that shear viscosity introduces an additional damping term in the tensor  equation, modifying both the transfer function and the  energy density power spectrum. For a constant shear viscosity-to-Hubble ratio the transfer function acquires an extra red tilt, while a time-dependent viscosity leads to a running spectral index $\Omega_\text{GW}\sim k^{n_\text{eff}(k)}$  controlled by the time evolution of the mean free path of the viscous fluid. Our analysis provides a general framework to analytically quantify how shear viscosity can alter the primordial gravitational wave background in standard and non-standard post-inflationary scenarios. As a case study we evaluate the effect of viscosity of the electron-photon-baryon plasma, on both the transfer function and the normalized energy density, finding a $k$-dependent blue tilt due to gravitational wave freeze-out from the viscous phase. This effect corresponds to a fractional difference of order $10^{-3}$. }


\begin{document}
\maketitle
\vskip 0.8 cm

\section{Introduction}

The detection of Gravitational Waves (GWs) by LIGO and Virgo~\cite{PhysRevLett.116.061102} has opened a new observational window on the Universe (see~\cite{Maggiore:2007ulw,Caprini:2018mtu} for reviews). Among the most promising targets for next-generation detectors are primordial Gravitational Waves (pGWs), which provide a unique probe of the physics of the very early Universe~\cite{LiGong:2024qmt}. They may originate from quantum fluctuations during inflation, from secondary (induced) scalar perturbations, or from non-inflationary sources such as topological defects and first-order phase transitions. Primordial signatures may also arise in alternative scenarios including bouncing cosmologies~\cite{Pinto-Neto:2017yez,Ito:2022lye,Ben-Dayan:2024aec, Conzinu:2024cwl, Conzinu:2025sot}. A vast literature explores the observable consequences of these mechanisms (see e.g.~\cite{Steinhardt:1993kv,Battye:1996pr,Bucher:1997xs,Krauss:2010ty,Tamayo:2015qla,Ricciardone:2016lym,Miron-Granese:2017qas,Graef:2018fzu,Wang:2019cvw,Bernal:2019lpc,Miroszewski:2020lsq,Ringwald:2020vei,Bari:2021xvf,Odintsov:2021urx,Odintsov:2022cbm,Yu:2023lmo,Zhumabek:2023wka,Dimopoulos:2023izi,Demir:2023jzy,AzarAgGhaleh:2024kff,Odintsov:2024sbo,Creminelli:2024qpu,Li:2024dce,Micheletti:2025cit,Khodadi:2025wuw,Kubo:2025jla,Caldwell:2022qsj}).

In standard cosmology, tensor modes generated during inflation evolve almost unaltered while they remain outside the horizon. After horizon re-entry, their propagation is governed by the dynamical properties of the cosmological medium, typically modeled as an ideal  fluid. In this case, the transfer function describing the sub-horizon evolution and the associated spectral energy density of pGW are well understood \cite{Caprini:2018mtu}. However, the perfect-fluid approximation is an idealization: whenever the primordial plasma is not in exact local equilibrium, dissipative phenomena arise and the cosmic medium could better described as an imperfect fluid with viscosity.

Shear viscosity provides a generic and physically motivated source of dissipation. It can arise from finite particle interaction rates, or more generally from any departure from instantaneous kinetic equilibrium. Its effect on GWs is conceptually similar to that of anisotropic stresses from free-streaming neutrinos~\cite{Weinberg:2004kg,Watanabe:2006qe}, which are known to be one of the main sources of suppression for the amplitude of sub-horizon tensor modes. However, while neutrino damping has been extensively analyzed, the explicit role of shear viscosity in modifying the tensor transfer function  has received comparatively little  attention \cite{Domenech:2025bvr}, despite its relevance in both standard and non-standard early-Universe scenarios.

In this work we provide a systematic analytical study of shear-viscous effects on the propagation of primordial gravitational waves after inflation, assuming instantaneous reheating. We show that shear viscosity introduces an effective friction term in the tensor mode equation, modifying both the transfer function and the spectral energy density. We derive exact solutions for a constant viscosity-to-Hubble ratio and develop a controlled perturbative treatment for time-dependent viscosity. This framework reveals how a transient viscous phase leaves a frozen suppression in the amplitude of sub-horizon modes once viscosity becomes negligible. Moreover,  we identify a viscous freeze-out mechanism by which viscous damping leaves a permanent imprint on sub-horizon tensor modes once they exit the hydrodynamic regime, introducing a $k$ dependent damping in $\Omega_{\text{GW}}$. 

Applying this framework to the photon-baryon-electron plasma, we obtain a realistic estimate of the resulting scale-dependent running of the tensor spectrum, finding a characteristic $k$-dependent blue tilt and a fractional suppression in $\Omega_{\rm GW}$ of order $10^{-3}$ in the LSS-CMB window. 

Our analytic results provide a fully controlled description of viscous gravitational-wave propagation that can be directly extended to non-standard dissipative scenarios such as warm-inflation reheating or strongly interacting hidden sectors, opening a new avenue to connect primordial gravitational-wave observations to microphysical properties of the early Universe.

The paper is organized as follows. In Sect.~\ref{sec2} we review the transfer function formalism in the presence of viscous damping. In Sect.~\ref{sec3} we study the case of constant viscosity-to-Hubble ratio, while Sect.~\ref{sec4} addresses a time-dependent viscosity. Sect.~\ref{sec5} describes the transfer function and the normalized energy density in radiation- and matter-dominated eras. In Sect.~\ref{sec7} we apply the framework to the photon-baryon plasma before recombination, finding as expected a $k$-dependent small suppression in $\Omega_\text{GW}$. Conclusions are drawn in Sect.~\ref{conclusion}.

\textit{Conventions}: we adopt units $c=\hbar=1$ and mostly-minus metric signature $(+,-,-,-)$. Greek indices denote space-time components, and $\lambda_p^2=M_p^{-2}=8\pi G$.

\section{The transfer function and the relative energy power spectrum with shear viscous damping}\label{sec2}

During the inflationary epoch, a stochastic background of GWs is generated by quantum vacuum fluctuations of the geometry, which become classical due to the evolution of the FLRW metric.  In these models, the perturbations
evolve until they cross the horizon, after which they freeze until their physical
wavelength becomes comparable with the Hubble radius and start to
evolve again in the standard cosmological epochs. This characteristic behavior 
leaves the gravitational waves unchanged by the subsequent phases
of evolution of the Universe, up to the re-entering time. Thanks to
this property, inflationary primordial GWs are an
extreme valuable source of information about the earliest time of
the Universe and can give us clues about physics beyond the Standard
Model of particle physics and to Quantum Gravity. In this section we present the basic formalism that is used in the paper, giving a brief review of the tensor transfer function and the modifications to the GWs propagation equation in the case of Starobinsky gravity or in the presence of background shear viscous fluids.

In general, the primordial spectrum generated during any inflationary epoch  at the horizon
crossing is given by:
\begin{equation}
\mathcal{P}_h(k)=\frac{2k^{3}}{\pi^{2}}|h_{k,\text{prim}}|^{2},\label{eq:1}
\end{equation}
where \( h_{k,\text{prim}} \) denotes the amplitude of gravitational-wave modes on super-horizon scales.\footnote{The sum over polarizations has already been taken into account.} After the horizon crossing the amplitude of GW is frozen and stay
unchanged until their re-entering.

To evaluate the energy density carried by pGWs we superimpose the metric perturbations
$h_{\mu\nu}$ on the background geometry as $g_{\mu\nu}=\bar{g}_{\mu\nu}+h_{\mu\nu}$
where $\bar{g}_{\mu\nu}=\text{diag}(1,-a^{2}(t),-a^{2}(t),-a^{2}(t))$  is
the background FLRW metric in cosmic time. The energy momentum tensor
of gravitational waves is given by (neglecting terms $O(h^3)$)
\begin{equation}
T_{\mu\nu}^{\text{GW}}=\frac{1}{4\lambda_{p}^{2}}\langle\nabla_{\mu}h_{\alpha\beta}\nabla_{\nu}h^{\alpha\beta}\rangle,\label{eq:2}
\end{equation}
where the covariant derivatives are done with respect to the background
metric $\bar{g}_{\mu\nu}$ and the angular parenthesis represents the expectation
value, that can be classical or quantum mechanical. The 00 component
of the stress energy tensor is interpreted as the energy density of
GWs
\begin{equation}
\rho_{h}(\tau)\equiv T_{00}^{\text{GW}}=\frac{1}{4\lambda_{p}^{2}}\langle\dot{h}_{ij}\dot{h}^{ij}\rangle=\frac{1}{4\lambda_{p}^{2}a^{2}}\langle h_{ij}^{\prime}h^{\prime}{}^{ij}\rangle,\label{eq:3}
\end{equation}
with $\dot{}\equiv \frac{\partial}{\partial t}$, the derivative with respect to the cosmic time, 
$^{\prime}\equiv\frac{\partial}{\partial \tau}$ the derivation with respect to the conformal time
$\tau$ defined as $a d\tau=dt$ and  $h_{\mu\nu}$ in the Transverse-Traceless (TT) gauge.  From Eq.~\eqref{eq:3} we have in terms of the Fourier modes
\begin{equation}
\rho_{h}(\tau)=\frac{1}{2\lambda_{p}^{2}a^{2}}\langle h_{+}^{\prime2}+h_{\times}^{\prime2}\rangle=\frac{1}{2\lambda_{p}^{2}a^{2}}\underset{A=+,\times}{\sum}\frac{1}{(2\pi)^{3}}\int d^{3}k\,|h_{A,k}^{\prime}(\tau)|^{2},\label{eq:4}
\end{equation}
where $h_+$ and $h_\times$ represent the usual polarization of gravitational waves, and we assumed that $h_{A}(\tau,\mathbf{x})$ is real hence its Fourier components satisfy $h_{A,-\mathbf{k}}(\tau)=h_{A,\mathbf{k}}^{*}(\tau)$
and $h_{A,\mathbf{k}}(\tau)=h_{A,k}(\tau)$, with comoving wavenumber $k=|\mathbf{k}|$.

Since we are interested in the evolution of the GWs when they re-enter
the horizon, let us introduce the \textit{Transfer
Function} $\mathcal{T}(k,\tau)$ as follows
\begin{equation}
h_{k}(\tau)=h_{k,\text{prim}}\mathcal{T}(k,\tau).\label{eq:10}
\end{equation}
Here $\mathcal{T}(k,\tau)$ encodes the time evolution of the pGWs after the horizon crossing that happens during
the inflationary period. Gravitons in the Bunch-Davies vacuum do not
have a preferred polarization, so we can assume $|h_{+,k}(\tau)|^{2}=|h_{\times,k}(\tau)|^{2}$.
Inserting Eq.~\eqref{eq:10} in Eq.~\eqref{eq:4} we arrive at
\begin{equation}
\rho_{h}(\tau)=\frac{1}{4\lambda_{p}^{2}a^{2}}\int d\ln\,k\,\mathcal{P}_h(k)\left[\mathcal{T}^{\prime}(k,\tau)\right]^{2}\label{eq:11}.
\end{equation}
It is customary to define the  \textit{normalized energy density power spectrum} as
\begin{equation}
\Omega_{\text{GW}}(k,\tau)=\frac{1}{\rho_{cr}(\tau)}\frac{d\rho_{h}(\tau)}{d\,\ln k},\label{eq:13}
\end{equation}
where $\rho_{cr}(\tau)=\frac{3}{a^{2}\lambda_{p}^{2}}\mathcal{H}^{2}$,
with $\mathcal{H}=\frac{a^{\prime}}{a}$. In this way it is possible
to write Eq.~\eqref{eq:13} as a function of the transfer function
\begin{equation}
\Omega_{\text{GW}}(k,\tau)=\frac{\mathcal{P}_h(k)}{12\,\mathcal{H}^{2}(\tau)}\left[\mathcal{T}^{\prime}(k,\tau)\right]^{2}.\label{eq:14}
\end{equation}

Since we are not interested in the shape of the pGWs power spectrum, whose information is encoded in $\mathcal{P}_h(k)$,
but rather in how the subsequent cosmological epochs (whether standard or non-standard) modify its evolution, we will focus on the evaluation
of the transfer function. In particular we will assume the possible effects of Starobinsky gravity or the presence of anisotropic contributions in the background sources.  The latter has been extensively studied in the context of neutrino's free streaming \cite{Weinberg:2004kg,Watanabe:2006qe}, while less attention has been devoted to shear viscosity \cite{Domenech:2025bvr}.
Remarkably both Starobinsky gravity and General Relativity with shear viscous background, generate the same mathematical modification in the gravitational wave propagation equation in terms of an additional friction term denoted with $\delta(\tau)$ \cite{Fanizza:2021ngq, Fanizza:2022pvx, Fanizza:2020hat, Conzinu:2025sot}
\begin{equation}
h_{k}^{\prime\prime}+2\mathcal{H\,}\left(1+\delta(\tau)\right)\,h_{k}^{\prime}+k^{2}\,h_{k}=0,\label{eq:15}
\end{equation}
where we neglected the polarization index to lighten the notation. The function $\delta(\tau)$ encodes the physical origin of the modification of the evolution equation. In the case of Starobinsky gravity \cite{Fanizza:2020hat}
\begin{equation}
    \delta(\tau)\equiv c_2 \mathcal{H}^{-1} \frac{dR}{d\tau}(1+2 c_2 R)^{-1},\label{starofric}
\end{equation} 
where $c_2$ is the dimensionful coefficient governing the magnitude of the square of the Ricci scalar in the gravitational action (i.e. $\mathcal{L}_{\text{GR}}\sim R+c_2R^2$). 

A similar effect is generated in the case of the presence of background viscous sources with energy-momentum tensor given by \cite{Landau:1987fm,Pavone:2025vru,Fanizza:2021ngq,Fanizza:2022pvx,Misner_Thorne_Wheeler_1973,Weinberg_1972,Gron_1990,Anand_Chaubal_Mazumdar_Mohanty_2017,Montanari_Venanzi_2017,Muronga_2004,Ganguly:2021pke} and references therein

\begin{equation}
\begin{aligned}
T^{\text{vis}}_{\alpha\beta}&=\rho u_{\alpha}u_{\beta}-\bar{p}\,\Delta_{\alpha\beta}+\pi_{\alpha\beta},\\ \label{eq:9}
\Delta_{\alpha\beta}&=g_{\alpha\beta}-u_\alpha u_\beta,\\
\pi_{\alpha\beta}&=2\eta\sigma_{\alpha\beta}=2\eta\left(\Delta_\alpha^{(\mu|}\Delta_\beta^{|\nu)}-\frac13\Delta_{\alpha\beta}\Delta^{\mu\nu}\right)\nabla_{\mu}u_\nu,\\
\bar{p}&=p-\zeta\nabla_{\mu}u^{\mu},
\end{aligned}
\end{equation}
where the round brackets denote symmetrization, $u^{\alpha}$ is the time-like fluid four-velocity ($u_{\alpha}u^{\alpha}=1$)
that we will assume to be comoving $u^{\alpha}=\delta_{0}^{\alpha}$,
$\rho$ is the energy density of the fluid, $\eta$ the shear
viscosity, $\bar{p}$
is the effective pressure and $\zeta$ the bulk viscosity. 
In this case the extra viscous term is \cite{Fanizza:2022pvx, Fanizza:2021ngq, Conzinu:2025sot}, 
\begin{equation}
    \delta(\tau)\equiv\frac{H_{V}(\tau)}{H(\tau)}\qquad\text{with}\qquad
    H_{V}\equiv\lambda_{p}^{2}\eta(\tau), \label{fricvis}
\end{equation}
where $H$ the Hubble parameter.  Upon the substitution given in Eq.~\eqref{eq:10}
we have that the transfer function evolves over time as
\begin{equation}
\begin{aligned}
\mathcal{T}^{\prime\prime}(k,\tau)+2\mathcal{H\,}&\left(1+\delta(\tau)\right)\,\mathcal{T}^{\prime}(k,\tau)+k^{2}\,\mathcal{T}(k,\tau)=0,\\\\
\text{with}\qquad\mathcal{T}(k,\tau_i)&=1,\qquad\text{and}\qquad\mathcal{T}^{\prime}(k,\tau_i)=0.
\end{aligned}\label{eq:15-1}
\end{equation}
The initial conditions are chosen to preserve the initial shape
of the pGWs spectrum $h_{k,\text{prim}}$ at the end of the inflationary period
 $\tau_i$. These are meant to be valid for modes that are super-horizon at the end of the inflationary period, i.e. such that $k\tau_i\ll1$.

\section{Spectral energy density with a  power law scale factor and constant \texorpdfstring{$\delta$}{delta}} \label{sec3}
First of all let us analyze the case of a power law scale factor given
by $a(\tau)\sim\tau^{\beta}$
and a constant $\delta(\tau)=\delta>0$. In this case Eq.~\eqref{eq:15-1}
can be solved as
\begin{equation}
\mathcal{T}_{\nu}(k,\tau)=\frac{1}{\left(k\tau\right)^{\nu-1}}\left[C_{1}\left(k\right)j_{\nu-1}\left(k\tau\right)+C_{2}\left(k\right)y_{\nu-1}\left(k\tau\right)\right],\label{eq:32-1}
\end{equation}
where $\nu=\beta(1+\delta)$ and $j_{\nu}(x)$ and $y_{\nu}(x)$ are
the spherical Bessel functions\footnote{For integer $\nu$ the definition for these functions is $j_\nu(x) = (-x)^\nu \left( \frac{1}{x} \frac{d}{dx} \right)^\nu \left(\frac{\sin x}{x}\right)$ and $y_\nu(x) = (-x)^\nu \left( \frac{1}{x} \frac{d}{dx} \right)^\nu \left(\frac{-\cos x}{x}\right)$. In a broader context where $\nu\notin \mathbb{N}$, we have $j_{\nu}(x)
= \sqrt{\frac{\pi}{2x}}\,
J_{\nu+1/2}(x) $ and $y_\nu=\sqrt{\frac{\pi}{2x}}\,Y_{\nu+1/2}(x)$, where $Y_\alpha(x)=\cot(\alpha\pi)\,J_\alpha(x)-\csc(\alpha\pi)\,J_{-\alpha}(x)$, and $J_\alpha(x)=\sum_{m=0}^{\infty}\frac{(-1)^m}{m!\,\Gamma(m+\alpha+1)}\left(\frac{x}{2}\right)^{2m+\alpha}$.}  of the first and second kind respectively and $C_1$ and $C_2$ are integrations constants.
Using the following relevant property of the Bessel functions
\begin{equation}
\frac{d}{dx}\left(\frac{z_{\nu}(x)}{x^{\nu}}\right)=-\frac{z_{\nu+1}(x)}{x^{\nu}},\label{eq:33-1}
\end{equation}
where $z_{\nu}(x)$ is a generic spherical Bessel function, we arrive at
\begin{equation}
\mathcal{T}_{\nu}^{\prime}(k,\tau)=-\frac{k}{\left(k\tau\right)^{\nu-1}}\left[C_{1}\left(k\right)j_{\nu}\left(k\tau\right)+C_{2}\left(k\right)y_{\nu}\left(k\tau\right)\right].\label{eq:34-1}
\end{equation}

Imposing the initial condition given in Eq.~\eqref{eq:15-1} and setting $\tau_i=0$, we have
\begin{equation}
\mathcal{T}_{\nu}(k,\tau)=\frac{\Gamma\left(\nu+\frac{1}{2}\right)2^{\nu}}{\left(k\tau\right)^{\nu-1}\sqrt{\pi}}j_{\nu-1}\left(k\tau\right),\label{eq:36-1}
\end{equation}
with $\Gamma(x)$ the Euler  gamma function. From Eq.~\eqref{eq:34-1}
we can conclude that
\begin{equation}
\mathcal{T}_{\nu}^{\prime}(k,\tau)=-\frac{\Gamma\left(\nu+\frac{1}{2}\right)2^{\nu}k}{\left(k\tau\right)^{\nu-1}\sqrt{\pi}}j_{\nu}\left(k\tau\right).\label{eq:37-1}
\end{equation}
For the sub-horizon modes $k\tau\gg1$, we
can use the following asymptotic relations for the $n$-th order spherical Bessel functions to determine the behavior
of the spectrum when the wavelength of the incoming GW varies
\begin{equation}
j_{\nu}(x)\overset{x\gg1}{\sim}\frac{1}{x}\sin\left(x-\nu\frac{\pi}{2}\right),\,\,\,\,\,\,\,\,\,\,\,\,\,\,
y_{\nu}(x)\overset{x\gg1}{\sim}-\frac{1}{x}\cos\left(x-\nu\frac{\pi}{2}\right).
\label{eq:38-1}
\end{equation}

We find that a constant ratio between the shear viscosity parameter
$H_{V}$ and $H$ introduces a red-tilt in the transfer
function for the sub-horizon modes ($k\tau\gg1$ and neglecting constant normalization factors)
\begin{equation}
\mathcal{T}_{\nu}(k,\tau)\sim \frac{k^{-\beta(1+\delta)}}{a(\tau)^{1+\delta}},\label{eq:18-1}
\end{equation}
and in the energy density power spectrum
\begin{equation}
\frac{\Omega_{\text{GW}}(k,\tau_{0})}{\mathcal{P}_h(k)}\sim \frac{k^{2-2\beta(1+\delta)}}{a^{2(1+\delta-1/\beta)}}.\label{eq:19-1}
\end{equation}

Here few comments are in order.  
The presence of a constant friction term $\delta$ (i.e. a constant ratio between the viscous parameter and the Hubble rate, 
\(\delta = H_{V}/H\) in the viscous case), produces a stronger damping of the sub-horizon tensor modes.  
Their amplitude no longer scales as \(a^{-1}\), as in the standard case, 
but as \(a^{-(1+\delta)}\). As a consequence, the associated energy density does not 
dilute as radiation, i.e. \(a^{-4}\), but more rapidly, following the scaling law 
\(a^{-(4+2\delta)}\).  This enhanced dilution translates into a red-tilt of the spectrum: 
the amplitude of high-frequency (large-\(k\)) modes is increasingly suppressed. 
In spectral terms, defining $n_0$ as the spectral tilt of the normalized energy density in the absence of viscous damping, we have  
\begin{equation}
\Omega_{\text{GW}}\propto k^{n_0+\Delta n},
\end{equation}
where $\Delta n$ encodes the deviation from the ideal case. In terms of the tensor power spectrum spectral index
\begin{equation}
    n_t\equiv\frac{d\ln\mathcal{P}(k)}{d\ln k},
\end{equation}
we have $n_0=n_t+2-2\beta$, which gives the standard $n_0=n_t$ for modes that re-enter in radiation, and $n_0=n_t-2$ for modes that re-enter during matter domination.
The additional friction shifts the energy density power spectrum spectral index by 
\(\Delta n= -2\beta\,\delta\), corresponding to 
\(\Delta n = -2\delta\) for modes entering during radiation domination and 
\(\Delta n = -4\delta\) for those entering during matter domination.
Since it is known that already the pGW signal from standard ``slow-roll'' inflation
should not be observable by next generation gravitational antennas such as LISA \cite{Auclair:2022jod}, ET \cite{Branchesi:2023mws}, DECIGO \cite{Kawamura:2020pcg}, we conclude that this reddening and additional dilution effect would suppress even more the signal at those frequency and sensitivities scales.

\section{Spectral energy density with a  power law scale factor and evolving \texorpdfstring{$\delta$}{delta}}
\label{sec4}
We will now solve Eq.~\eqref{eq:15} assuming an evolving $\delta(\tau)$ at leading order in a $\delta$
expansion. In order to be consistent we have to require $\delta(\tau)\ll1$,
which implies in Starobinsky gravity $c_2\frac{dR}{d\tau}\ll(1+2c_2 R)\mathcal{H}$ or  $H_{V}(\tau)\ll H$ in the viscous case. The last condition is generically fulfilled by viscous modifications both in early-time  (as presented in Sect.~\ref{sec7}), and in late-time cosmology. In fact the upper  bound given by the estimation of the shear viscosity today
given by the cross section per unit mass $\sigma/m$ from Abell 3827 \cite{Massey:2015dkw}
gives $H_{V}(\tau_{0})\approx5\times10^{-3}H_{0}$, where $\tau_{0}$
is the conformal time today. 
We are interested in an analytical approximation
for the transfer function to highlight possible discrepancies in the
overall functional behavior of the normalized  spectral energy
density.

Under these assumptions Eq.~\eqref{eq:15-1}
can be rewritten as follows
\begin{equation}
\Big(\hat{L}_{\beta,\tau}^{(0)}+\delta(\tau)\hat{L}_{\beta,\tau}^{(1)} \Big)\,\mathcal{T}_{\beta}(k,\tau)=0\qquad,\qquad\mathcal{T}_{\beta}(k,\tau_i)=1\qquad,\qquad \mathcal{T}_{\beta}^{\prime}(k,\tau_i)=0\,,\label{eq:21}
\end{equation}
where we have defined the following differential operators
\begin{equation}
\hat{L}_{\beta,\tau}^{(0)} \equiv \frac{\partial^{2}}{\partial\tau^{2}}+\frac{2\beta}{\tau}\frac{\partial}{\partial\tau}+k^{2},\qquad\qquad
\hat{L}_{\beta,\tau}^{(1)} \equiv \frac{2\beta}{\tau}\frac{\partial}{\partial\tau}.
\label{eq:22}
\end{equation}
In order to find our desired analytic approximation, we assume that
the transfer function can be written as follows
\begin{equation}
\mathcal{T}_{\beta}(k,\tau)=\mathcal{T}_{\beta}^{(0)}(k,\tau)+\mathcal{T}_{\beta}^{(1)}(k,\tau)+O(\delta^{2})\,,\label{eq:23}
\end{equation}
where $\mathcal{T}_{\beta}^{(1)}$ is treated as a first order correction to
$T_{\beta}^{(0)}$ in the $\delta$ expansion. Then, inserting the
last equality into Eq.~\eqref{eq:14}, we get the following.
\begin{equation}
\frac{\Omega_{\text{GW}}(k,\tau)}{\mathcal{P}_h(k)}=\frac{\mathcal{T}_{\beta}^{(0)\prime}(k,\tau)^{2}}{12\,\mathcal{H}^{2}(\tau)}+\frac{\mathcal{T}_{\beta}^{(0)\prime}(k,\tau)\,\,\mathcal{T}_{\beta}^{(1)\prime}(k,\tau)}{6\,\mathcal{H}^{2}(\tau)}+O(\delta^{2}).\label{eq:24}
\end{equation}

In order to take full advantage of Eq.~\eqref{eq:24}, we need to find
the explicit expression for $T_{\beta}^{(1)}$. To this end, we insert
Eq.~\eqref{eq:23} into Eqs.~\eqref{eq:21} and keep only terms up to
$O(\delta)$. As a result, the leading order evolution equation is
\begin{equation}
\hat{L}_{\beta,\tau}^{(0)}\mathcal{T}_{\beta}^{(0)}(k,\tau)=0,\qquad\qquad
\mathcal{T}_{\beta}^{(0)}(k,\tau_i)=1,\qquad\qquad
\mathcal{T}_{\beta}^{(0)\prime}(k,\tau_i)=0\,,
\label{eq:25}
\end{equation}
whereas the first order correction must satisfy
\begin{equation}
\hat{L}_{\beta,\tau}^{(0)}\left[\mathcal{T}_{\beta}^{(1)}(k,\tau)\right]=-\delta\hat{L}_{\beta,\tau}^{(1)}\mathcal{T}_{\beta}^{(0)}(k,\tau),\qquad\qquad
\mathcal{T}_{\beta}^{(1)}(k,\tau_i)=0,\qquad\qquad
\mathcal{T}_{\beta}^{(1)\prime}(k,\tau_i)=0,\label{eq:26}
\end{equation}
where we neglected second order term and we imposed the initial conditions according to Eq.~\eqref{eq:21}.
\\ \\
\textbf{The leading order solution}\\
Eq.~\eqref{eq:25} at the lowest order can be solved directly
using the results of the previous subsection upon the substitution
$\nu\to\beta$ ($\tau_i=0$)
\begin{equation}
\mathcal{T}_{\beta}^{(0)}(k,\tau)=\frac{\mathcal{A}_\beta}{\left(k\tau\right)^{\beta-1}}j_{\beta-1}\left(k\tau\right),\qquad\qquad \mathcal{T}_{\beta}^{(0)\prime}(k,\tau)=-\frac{\mathcal{A}_\beta }{\left(k\tau\right)^{\beta-1}}k\,j_{\beta}\left(k\tau\right)\,,\label{eq:36}
\end{equation}
where we defined
\begin{equation}
\mathcal{A}_\beta\equiv2^\beta\frac{\Gamma(\beta+\frac12) }{\sqrt\pi}\,.
\end{equation}
Using Eq.~\eqref{eq:24}, at the leading order in $k$ and 0-th order
in $\delta$, we have, as we have already stated in the previous section
\begin{equation}
\frac{\Omega_{\text{GW}}^{(0)}(k,\tau)}{\mathcal{P}_h(k)}\simeq\begin{cases}
\begin{array}{cccccc}
const\,, &   & \text{Radiation} \quad(\beta=1)\,,\\
\\
k^{-2}\,, &   & \text{Matter} \quad (\beta=2)\,,
\end{array}
\end{cases}\label{eq:39}
\end{equation}
in concordance with \cite{Caprini:2018mtu}.
\\
\\
\textbf{The 1st order solution}\\
Once we have obtained the leading order transfer function, we are now
able to solve Eq.~\eqref{eq:26}. To this end we introduce the Green function $G(\tau,\tau')$, so the solution can be  written as
\begin{equation}
\mathcal{T}_{\beta}^{(1)}(k,\tau)=-\int_{\tau_i}^{\tau}d\tau'\,G\left(\tau,\tau'\right)\delta(\tau^{\prime})\hat{L}_{\beta,\tau'}^{(1)}\mathcal{T}_{\beta}^{(0)}(k,\tau'),\label{eq:40}
\end{equation}
with $G\left(\tau,\tau'\right)$ satisfying
\begin{equation}
\hat{L}_{\beta,\tau}^{(0)}G\left(\tau,\tau'\right)=\delta_{\text{Dirac}}\left(\tau-\tau'\right),\quad\quad
G\left(\tau,\tau'\right)\big|_{\tau=\tau_i}=0,\quad\quad
\frac{\partial}{\partial\tau}G(\tau,\tau')\big|_{\tau=\tau_i}=0\,,\label{eq:41}
\end{equation}
for all $\tau,\tau'\in\mathbb{R}^+$. The solution is given by
\begin{equation}
G\left(\tau,\tau'\right)=-\Theta\left(\tau-\tau'\right)\frac{\left(k\tau^{\prime}\right)^{\beta+1}}{k\left(k\tau\right)^{\beta-1}}\left[j_{\beta-1}\left(k\tau\right)y_{\beta-1}\left(k\tau'\right)-y_{\beta-1}\left(k\tau\right)j_{\beta-1}\left(k\tau'\right)\right]\,,\label{eq:30}
\end{equation}
where $\Theta(x)$ is the Heaviside function. Inserting this result in Eq.~\eqref{eq:40} and  the second
definition of Eq.~\eqref{eq:22}, the first order correction of the transfer
function is 
\begin{equation}
\begin{aligned}
\mathcal{T}_{\beta}^{(1)}(k,\tau)=&-\mathcal{A}_\beta\frac{2\beta k^{2}}{\left(k\tau\right)^{\beta-1}}\int_{0}^{\tau}d\tau'\,\delta(\tau')\,\tau'\bigg[j_{\beta-1}\left(k\tau\right)y_{\beta-1}\left(k\tau'\right)\\
&-y_{\beta-1}\left(k\tau\right)j_{\beta-1}\left(k\tau'\right)\bigg]j_{\beta}\left(k\tau'\right),
\end{aligned}\label{eq:53}
\end{equation}
whose derivative is given by
\begin{equation}
\begin{aligned}
\mathcal{T}_{\beta}^{(1)'}(k,\tau)=&\mathcal{A}_\beta\frac{2\beta k^{3}}{\left(k\tau\right)^{\beta-1}}\int_{0}^{\tau}d\tau'\,\delta(\tau')\,\tau'\bigg[j_{\beta}\left(k\tau\right)y_{\beta-1}\left(k\tau'\right)\\
&-y_{\beta}\left(k\tau\right)j_{\beta-1}\left(k\tau'\right)\bigg]j_{\beta}\left(k\tau'\right)\,.
\end{aligned}\label{eq:54}
\end{equation}

We can write the relative spectral energy density at a given conformal
time from Eq.~\eqref{eq:24} as
\begin{align}
\frac{\Omega_{\text{GW}}(k,\tau)}{\mathcal{P}_h(k)}=&\frac{\mathcal{A}_{\beta}^{2}\,k^{2}}{12\mathcal{H}\left(\tau\right)^{2}\left(k\tau\right)^{2\beta-2}}\Bigg\{j_{\beta}\left(k\tau\right)^{2}\label{eq:56}\\
&-4\beta j_{\beta}\left(k\tau\right)k^{2}\int_{0}^{\tau}d\tau'\,\tau'\,\delta(\tau')\,j_{\beta}\left(k\tau'\right)\Big[j_{\beta}\left(k\tau\right)y_{\beta-1}\left(k\tau'\right)-y_{\beta}\left(k\tau\right)j_{\beta-1}\left(k\tau'\right)\Big]\Bigg\}\,.\nonumber
\end{align}
For later convenience we introduce the following decomposition for the normalized energy density power spectrum
\begin{equation}
    \frac{\Omega_{\text{GW}}(k,\tau)}{\mathcal{P}_h(k)}=\frac{\Omega^{(0)}_{\text{GW}}(k,\tau)+\Omega^{(1)}_{\text{GW}}(k,\tau)}{\mathcal{P}_h(k)},
\end{equation}
where $\Omega^{(1)}_{\text{GW}}(k,\tau)$ is the linear correction in the $\delta$ expansion. This is the starting point of all the following derivations.

\section{Shear viscous damping in  standard cosmological epochs}\label{sec5}
In this section we apply the  treatment presented in the previous section in two relevant science cases, namely a radiation and matter dominated Universe with a power law friction term.  From now on, we will consider the friction term as arising from viscous background sources in General Relativity as in Eq.~\eqref{fricvis}, but we remind the reader that these results are also valid in Starobinsky gravity up to the identification of the friction parameter $\delta$ with the higher-curvature correction of Eq.~\eqref{starofric}. 

The description of the matter sector adopted follows an effective hydrodynamic approach, where dissipative effects are encoded in macroscopic transport coefficients such as the shear viscosity. This framework is meaningful only when microscopic interactions equilibrate the system over a characteristic mean-free-path $\lambda_{\mathrm{mfp}}$, such  that the medium behaves as a continuous fluid on  scales grater than $\lambda_{\text{mfp}}$. Moreover we stress that hydrodynamical validity imposes that the effective fluid description is valid for $\lambda_{\text{mfp}}\lesssim L_{\text{macro}}$, where $L_{\text{macro}}$ is the typical scale of the system, which is $L_{\text{macro}}\simeq H^{-1}$. This means that the mean-free-path cannot exceed  the Hubble radius, namely 
\begin{equation}
    \lambda_{\text{mfp}}\lesssim H^{-1}\,. \label{hydro1}
\end{equation}
This point has been thoroughly discussed in \cite{Ganguly:2021pke}, where it has been shown from kinetic theory argument that an effective shear viscous fluid description is meaningful. From kinetic theory arguments we can write the shear viscosity for a  relativistic fluid $X$ as \cite{Weinberg:1971mx, Pajer:2012qep,Domenech:2025bvr}
\begin{equation}
    \eta=\frac{4}{15}(\rho_X+p_X)t_{\text{relax}},\label{visgen}
\end{equation}
where $t_{\text{relax}}=\lambda_{\text{mfp}}/v_{\text{rel}}$, and $v_{\text{rel}}$ is the typical relative speed of the two interacting species. One would therefore expect that in the limit $\lambda_{\text{mfp}}\to\infty$, the shear viscosity becomes increasingly large. However this limit  does not correspond to a diverging viscosity, but rather to a free-streaming fluid, where hydrodynamical arguments no longer  apply and $\eta\to0$. Accordingly, the viscous description applies only within this hydrodynamic regime, i.e. for mean-free-path smaller than the Hubble radius and for perturbations whose physical wavelength is  larger than the microscopic equilibration scale. More explicitly we have 
\begin{equation}
k_{\mathrm{phys}} \lesssim\lambda_{\mathrm{mfp}}^{-1}, \qquad k_{\mathrm{phys}} = \frac{k}{a},\label{hydrolim}
\end{equation}
where $\lambda_{\text{mfp}}$ is in general a function of the scale factor.   Microscopically, the mean free path can be written as 
\begin{equation}
\lambda_{\mathrm{mfp}} = \frac{1}{\sigma\, n\,},
\end{equation}
where $\sigma$ is an appropriate scattering cross section and $n$ the number density of the scattering particles (e.g. electrons in the photon-electron-baryon plasma).

As the Universe expands, both $n$ and $\sigma$ may vary with time, causing $\lambda_{\mathrm{mfp}}=\lambda_{\text{mfp}}(\tau)$ to evolve and thereby modifying the range of scales over which the fluid description remains valid. Modes violating the condition given by the Eq.~\eqref{hydrolim} probe sub-mean-free-path physics, where  a kinetic-theory treatment would be required instead. Hence to model this viscous freeze-out for the modes that are smaller than the mean-free-path at a given time $\tau$, we define
\begin{equation}
k_{\text{vis}}(\tau)\equiv\lambda_{\text{mfp}}^{-1}(\tau)a,
\end{equation}
such that the viscosity is active only  for modes with $k\lesssim k_{\text{vis}}(\tau)$. We will assume  from now on (as in the case of the electron-photon-baryon plasma in Sect.~\ref{sec7}) that $k_{\text{vis}}(\tau)$ is a decreasing function of the scale factor. This allows us to introduce a viscous freeze-out time per mode $k$, defined implicitly by inverting $k_{\text{vis}}(\tau)=k$, to obtain $\tau_{\text{vis}}^{\text{exit}}(k)$ which gives the conformal time after which the mode $k$ exceeds the mean-free-path.

Since we assumed a decreasing behavior for the viscous comoving wavenumber we can also conclude that the larger  the $k$ the smaller $\tau_{\text{vis}}^{\text{exit}}$, hence high frequency modes are less affected by viscous damping. This already suggests that the resulting energy density spectrum in the presence of shear viscosity will display a (possibly running) blue-tilt with respect to the non-viscous case\footnote{Conversely if one assumes $k_{\text{vis}}(\tau)$ to be an increasing function of the scale factor the opposite could be concluded, hence smaller $k$ freeze-out earlier, so high frequency modes would be damped more, giving rise to a red-tilt.}. Finally,   modes that initially evolve within  the  viscous background eventually stop to feel the effect of viscosity for $\tau\gtrsim\tau_{\text{vis}}^{\text{exit}}(k)$. The last time scale relevant for the problem is the one that can be derived from Eq.~\eqref{hydro1} and defines the limit of validity for the hydrodynamical treatment regardless of the mode $k$ under exam. We call this conformal time $\tau_{\text{max}}$ after which $\lambda_{\text{mfp}}\gtrsim H^{-1}$ and the viscosity vanishes.
For the purpose of this work and as a natural way to parametrize shear viscosity,  we model the friction term as
\begin{equation}
\delta(\tau)\equiv\delta_{\text{max}}\left(\frac{\tau}{\tau_{\text{max}}}\right)^\alpha\Theta(\tau^{\text{exit}}(k)-\tau),\qquad\qquad\tau^{\text{exit}}(k)\equiv\min(\tau_{\text{max}},\tau_{\text{vis}}^{\text{exit}}(k)),\label{friction}
\end{equation}
 where $\Theta$ is the Heaviside function, $\delta_{\text{max}}>0$ is the maximum value for $\delta(\tau)$,  $\delta_{\text{max}}\ll$1 and $\alpha>0$ in order to preserve
the condition $\delta(\tau)\ll1$ and $\tau_{\text{max}}$ is the conformal
time where  the friction contribution vanishes for all the modes. 

All the treatment presented is focused on the amplitude of the transfer function, or the normalized energy density power spectrum thereof, hence we are not interested in eventual phase shifts introduced on the transfer function. We work in the the sub-horizon regime, so at the leading order in the $k\tau$ expansion.

\subsection{Power law friction in a radiation dominated Universe}
Let us discuss the above mentioned results for the transfer function and the normalized energy density power spectrum in a radiation
dominated Universe described by the friction term given in Eq.~\eqref{friction} 

For a radiation dominated Universe $\beta=1$ and $\tau_i=0$, we
have that the first order correction to Eq.~\eqref{eq:53}, after the substitution $\xi=k\tau'$
and for $\tau\leq\tau_{\text{max}}$, is
\begin{equation}
\mathcal{T}_{1}^{(1)}(k,\tau)=-\frac{2}{(k\tau_{\text{max}})^{\alpha}}\delta_{\text{max}}\int_{0}^{k\tau^{\star}}d\xi\,\xi^{\alpha+1}\bigg[j_{0}\left(k\tau\right)y_{0}\left(\xi\right)-y_{0}\left(k\tau\right)j_{0}\left(\xi\right)\bigg]j_{1}\left(\xi\right)\,,
\label{eq:57}
\end{equation}
where we defined $\tau^\star\equiv \min\{\tau^{\text{exit}}(k),\tau\}$.
By evaluating the transfer function at $\tau<\tau_{\text{max}}$ and taking
the sub-horizon limit $k\tau^\star\gg1$, the full transfer function\footnote{We keep only leading order terms in $\xi^\alpha$ and we average out the oscillatory parts $\cos^2\xi=\sin^2\xi=1/2$ and $\sin\xi\cos\xi=0$.} becomes
\begin{equation}
\mathcal{T}_{1}(k,\tau)\simeq\frac{\sin(k\tau)}{k\tau}\left(1-\frac{\delta_{\text{max}}}{\alpha}\left(\frac{\tau^\star}{\tau_{\text{max}}}\right)^{\alpha}\right)=j_{0}\left(k\tau\right)\left(1-\frac{\delta(\tau^\star)}{\alpha}\right)\,.\label{eq:57a}
\end{equation}
It is straightforward to check that the fractional difference $\Delta_\mathcal{T}$  is given
by
\begin{equation}
\Delta_\mathcal{T}\equiv\frac{\mathcal{T}_{1}^{(0)}(k,\tau)-\mathcal{T}_{1}(k,\tau)}{\mathcal{T}_{1}^{(0)}(k,\tau)}\simeq\frac{\delta(\tau^\star)}{\alpha}\,.\label{eq:57b}
\end{equation}

Similar calculations lead to the following asymptotic relation for
the normalized energy density power spectrum 
\begin{equation}
\frac{\Omega_{\text{GW}}(k,\tau)}{\mathcal{P}_h(k)}\simeq\frac{\alpha-2\delta(\tau^\star)}{12\alpha}\cos^{2}(k\tau),\label{eq:57c}
\end{equation}
so we can evaluate the fractional difference $\Delta_\Omega$  as
\begin{equation}
\Delta_\Omega\equiv\frac{\Omega^{(0)}_\text{GW}(k,\tau)-\Omega_{\text{GW}}(k,\tau)}{\Omega^{(0)}_\text{GW}(k,\tau)}\simeq\frac{2\delta(\tau^\star)}{\alpha}.\label{eq:57d}
\end{equation}
For all the modes that re-entered the horizon during this era the damping
of the spectrum is preserved throughout the following phases as we are going to show. All
the modes with $k\gg\tau_{\text{max}}^{-1}$ are inside the horizon during the
viscous phase, and for the modes inside the horizon the transfer function
scales as $1/a$ for $\tau>\tau_{\text{max}}$. In fact, assuming that $\tau_{\text{max}}\leq\tau_{\text{eq}}$, where $\tau_{\text{eq}}$ is the conformal time at radiation-matter cosmological transition, we have
that during the subsequent phases of
evolution (non-viscous radiation phase and matter domination) the modes of interest $k\gg\tau_{\text{max}}^{-1}$ are sub-horizon. In this case, following \cite{Fanizza:2022pvx}, we have that Eq.~\eqref{eq:15} can be diagonalized using the Mukhanov-Sasaki-like variable 
\begin{equation}
    v_k(\tau)=\mathcal{T}(k,\tau)a(\tau).
\end{equation}
Substituting the latter into Eq.~\eqref{eq:15}, we have
\begin{equation}
    v''_k+\left(k^2-\frac{ a''}{ a}\right)v_k=0, 
\end{equation}
hence for sub-horizon modes $k^2\gg a''/a$, it can be solved as 
\begin{equation}
    v_k(\tau)=A \cos(k\tau)+B\sin(k\tau),
\end{equation}
where $A$ and $B$ are integration constants fixed by the continuity of $\mathcal{T}(k,\tau)$ and its derivative at $\tau=\tau_{\text{max}}$ with the solution Eq.~\eqref{eq:57a}.

At the leading order in $(k \tau_{\text{max}})^{-1}$ and $\delta$ we have that the transfer function is 
\begin{equation}
\mathcal{T}(k,\tau>\tau_{\text{max}})\simeq\mathcal{T}_{1}(k,\tau_{\text{max}})\frac{\sin(k\tau)}{\sin(k\tau_\text{max})}\frac{a(\tau_{\text{max}})}{a(\tau)}\,.\label{eq:58}
\end{equation}
Indeed the fractional difference in the transfer function for $\tau>\tau_\text{max}$ is frozen
and takes its maximum value
\begin{equation}
\Delta_\mathcal{T} \simeq \frac{\delta(\tau^{\text{exit}}(k))}{\alpha}\qquad,\qquad{\tau>\tau_{\text{max}}}\,.\label{eq:59}
\end{equation}
\begin{figure}[ht!]
    \centering
    \begin{subfigure}{0.48\textwidth}
        \centering
       \includegraphics[height=4.5cm]{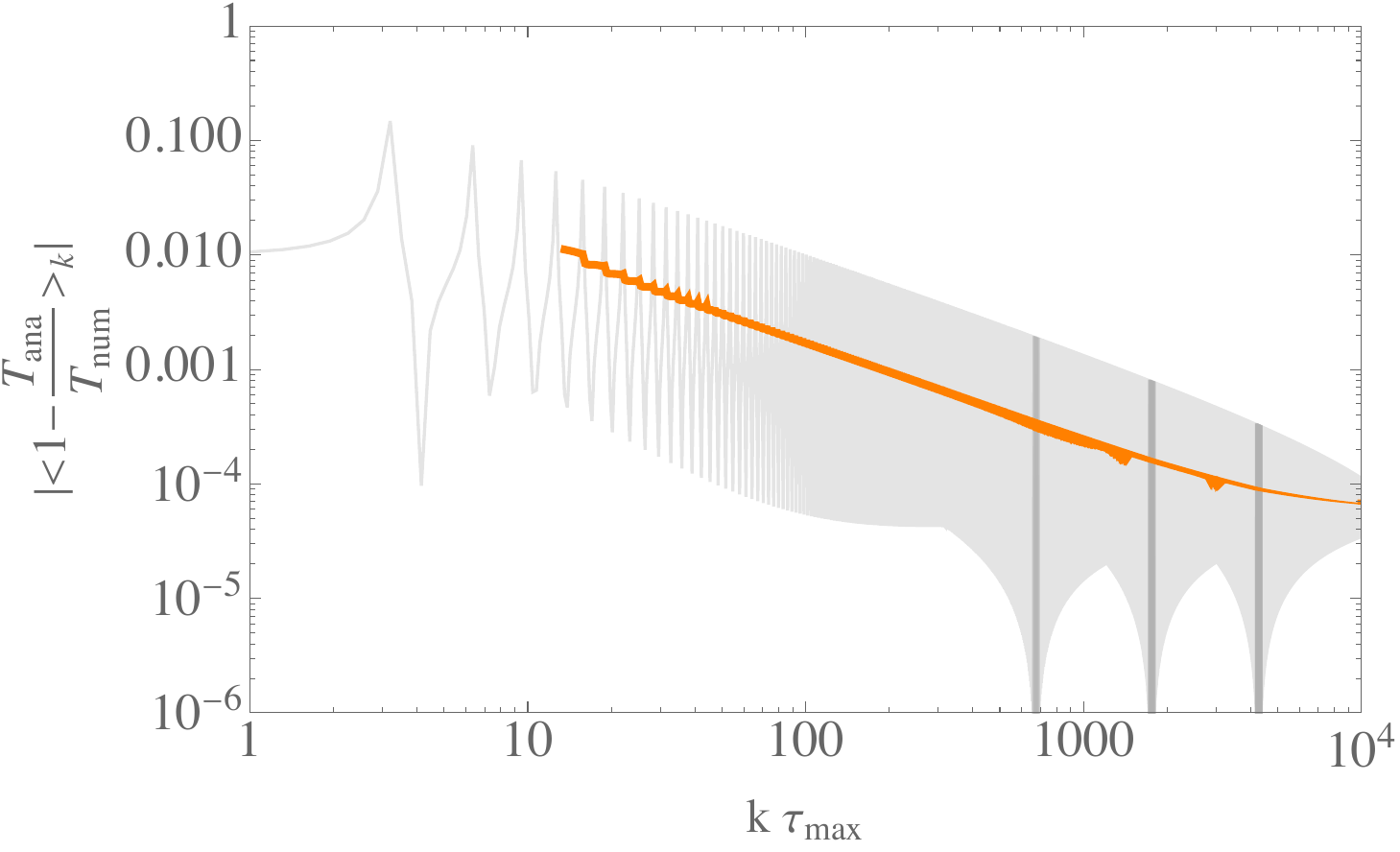}
       \caption{}
    \end{subfigure}
    \hfill
    \begin{subfigure}{0.48\textwidth}
        \centering
       \includegraphics[height=4.5cm]{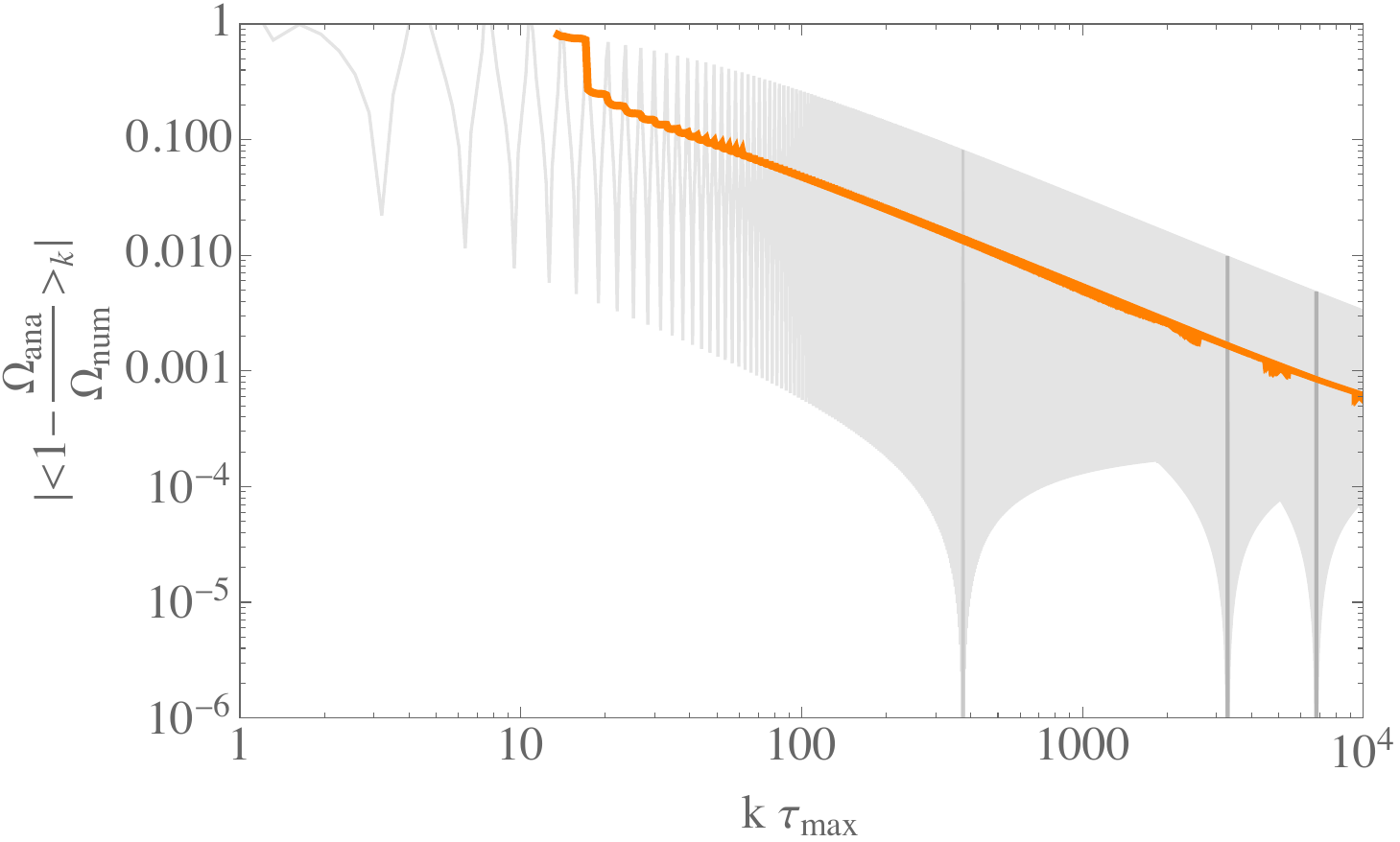}
       \caption{}
    \end{subfigure}
    \caption{Relative difference of the analytic and numeric transfer function and $\Omega_{\text{GW}}$ in a viscous radiation epoch evaluated at $\tau=\tau_{\text{max}}$, for $\alpha=1$, $\delta_{\text{max}}=10^{-2}$ and $\lambda_{\text{mfp}}=0$. In gray the relative difference, in orange the $k$ average.}\label{fig:rad}
\end{figure}
The same argument can be applied to the normalized energy density
power spectrum with the addition of the approximation $\mathcal{T}^{\prime}(k,\tau)\sim k\mathcal{T}(k,\tau)$
valid only for sub-horizon modes. So from Eqs.~\eqref{eq:57d}) and \eqref{eq:58}
and the definition of the spectral energy density we have
\begin{equation}
\Delta_\Omega\simeq\frac{2\delta(\tau^{\text{exit}}(k))}{\alpha}\qquad,\qquad\tau>\tau_{\text{max}}\,.\label{eq:59a}
\end{equation}
Hence also the spectral energy density is reduced by a $k$ dependent factor which depends only on the value of the friction term at the viscous freeze-out per mode $k$  and the power law coefficient. 

From Eq.~\eqref{eq:59a} we can also derive the effective tensor index and its running. Defining as usual the tensor spectral index as $n_t\equiv\frac{d\ln\mathcal{P}(k)}{d\ln k}$, for modes that re-enter in a radiation epoch we have $\Omega_{\text{GW}}^{(0)}\propto k^{n_0}$, with $n_0=n_t$. Taking into account the viscous contribution we find that defining
\begin{equation}
n_{\text{eff}}(k)\equiv\frac{d\ln\Omega_{\text{GW}}}{d\ln k}\,,\label{efftens}
\end{equation}
we have 
\begin{equation}
\begin{aligned}
n_{\text{eff}}(k)&=n_0+\Delta n(k),\\
\Delta n(k)&=-2\alpha\frac{\delta(\tau^{\text{exit}}(k))}{\alpha-2\delta(\tau^{\text{exit}}(k))}\frac{d\ln\tau^{\text{exit}}(k)}{d\ln k}\simeq-2\delta(\tau^{\text{exit}}(k))\frac{d\ln\tau^{\text{exit}}(k)}{d\ln k},
\end{aligned}
\end{equation}
where in the last equality we used the limit $\delta(\tau)\ll1$. Since $\tau^\text{exit}(k)$ is a decreasing function of $k$, $\Delta n(k)>0$, hence the spectrum acquires a \textit{blue-tilt} with respect to the non-viscous case. We stress here that $n_0$ itself can have a $k$ dependence since we are not assuming a particular inflationary model, and all the information of the cosmological epochs preceding the  radiation phase are encoded in it, since a radiation phase leaves the primordial spectral tilt unaltered.

Finally in the case one considers modes with $k/a\,\lambda_{\text{mfp}}\to0$, we have that $\tau^{\text{exit}}(k)=\tau_{\text{max}}$, hence both the transfer function and the spectral energy density are damped by a constant factor given by the $\delta_{\text{max}}/\alpha$ and $2\delta_{\text{max}}/\alpha$ for all modes.  To benchmark the accuracy of our approximation, in Fig.~\ref{fig:rad} we show the $k$-averaged relative difference between the analytic approximation Eqs.~\eqref{eq:57a} and \eqref{eq:57c} with the numerical result for the transfer function and energy density, at the end of the viscous era $\tau=\tau_{\text{max}}$, with a linear viscous scaling. In the deep sub-horizon regime, we have that the relative difference between the numerical solution and the analytic approximation of the transfer function is $O(10^{-3})$ at $k\tau_{\text{max}}\simeq100$. For $\Omega_{\text{GW}}$ the relative difference scales as $(k\tau_{\text{max}})^{-1}$.

\subsection{Power law friction in a matter dominated Universe}\label{sec6}
Let us now discuss the case of a matter dominated Universe, assuming the same power-law behaviour for the shear viscosity.
For a matter dominated background we have $\beta=2$. Substituting the above $\delta(\tau)$ Eq.~\eqref{friction} into Eq.~(\ref{eq:53}) and changing variable to $\xi=k\tau'$, for $\tau\leq\tau_{\text{max}}$ we obtain
\begin{equation}
\mathcal{T}_{2}^{(1)}(k,\tau)
=-\frac{12}{(k\tau)(k\tau_{\text{max}})^{\alpha}}\delta_{\text{max}}
\int_{0}^{k\tau^{\star}}d\xi\,\xi^{\alpha+1}\bigg[j_{1}\left(k\tau\right)y_{1}\left(\xi\right)-y_{1}\left(k\tau\right)j_{1}\left(\xi\right)\bigg]j_{2}\left(\xi\right)\,.
\label{eq:60}
\end{equation}
By evaluating the integral in the sub-horizon limit $k\tau^\star\gg1$, we can use the large argument approximations for the Bessel functions in Eqs.~\eqref{eq:38-1}. Under these conditions, the leading contribution to Eq.~(\ref{eq:60}) can be explicitly computed, giving
\begin{equation}
\mathcal{T}_{2}^{(1)}(k,\tau)\simeq
-\frac{6\,\delta_{\text{max}}}{\alpha}\left(\frac{\tau^\star}{\tau_{\text{max}}}\right)^{\alpha}
\frac{j_{1}(k\tau)}{k\tau}.
\label{eq:61}
\end{equation}
The full transfer function, including the first order correction in $\delta$, can thus be written as
\begin{equation}
\mathcal{T}_{2}(k,\tau)\simeq
3\,\frac{j_{1}(k\tau)}{k\tau}\left(1-\frac{2\delta(\tau^\star)}{\alpha}\right),
\label{eq:62}
\end{equation}
where we used the zeroth-order matter-dominated solution
$\mathcal{T}_{2}^{(0)}(k,\tau)=3j_{1}(k\tau)/(k\tau)$.

The fractional difference  for $\tau\leq\tau_{\text{max}}$ between the viscous and inviscid transfer functions is therefore
\begin{equation}
\Delta_{\mathcal{T}}\simeq\frac{2\delta(\tau^\star)}{\alpha}.
\label{eq:63}
\end{equation}
Similar calculations lead to the following asymptotic relation for the normalized spectral energy density
\begin{equation}
\frac{\Omega_{\text{GW}}(k,\tau)}{\mathcal{P}_h(k)}\simeq
\left(\frac{3}{16}-\frac{3\delta(\tau^\star)}{4\alpha}\right)\,\frac{\sin^{2}(k\tau)}{(k\tau)^{2}},
\label{eq:64}
\end{equation}
which reduces to the standard result $\Omega_{\text{GW}}^{(0)}/\mathcal{P}_h\sim 3/16(k^2\tau^2)^{-1}\sin^2(k\tau)$ in the limit $\delta\to0$.
The corresponding relative deviation reads
\begin{equation}
\Delta_\Omega \simeq\frac{4\,\delta(\tau^\star)}{\alpha}.
\label{eq:65}
\end{equation}
\begin{figure}[t]
    \centering
    \begin{subfigure}[b]{0.48\textwidth}
        \centering
       \includegraphics[height=4.4cm]{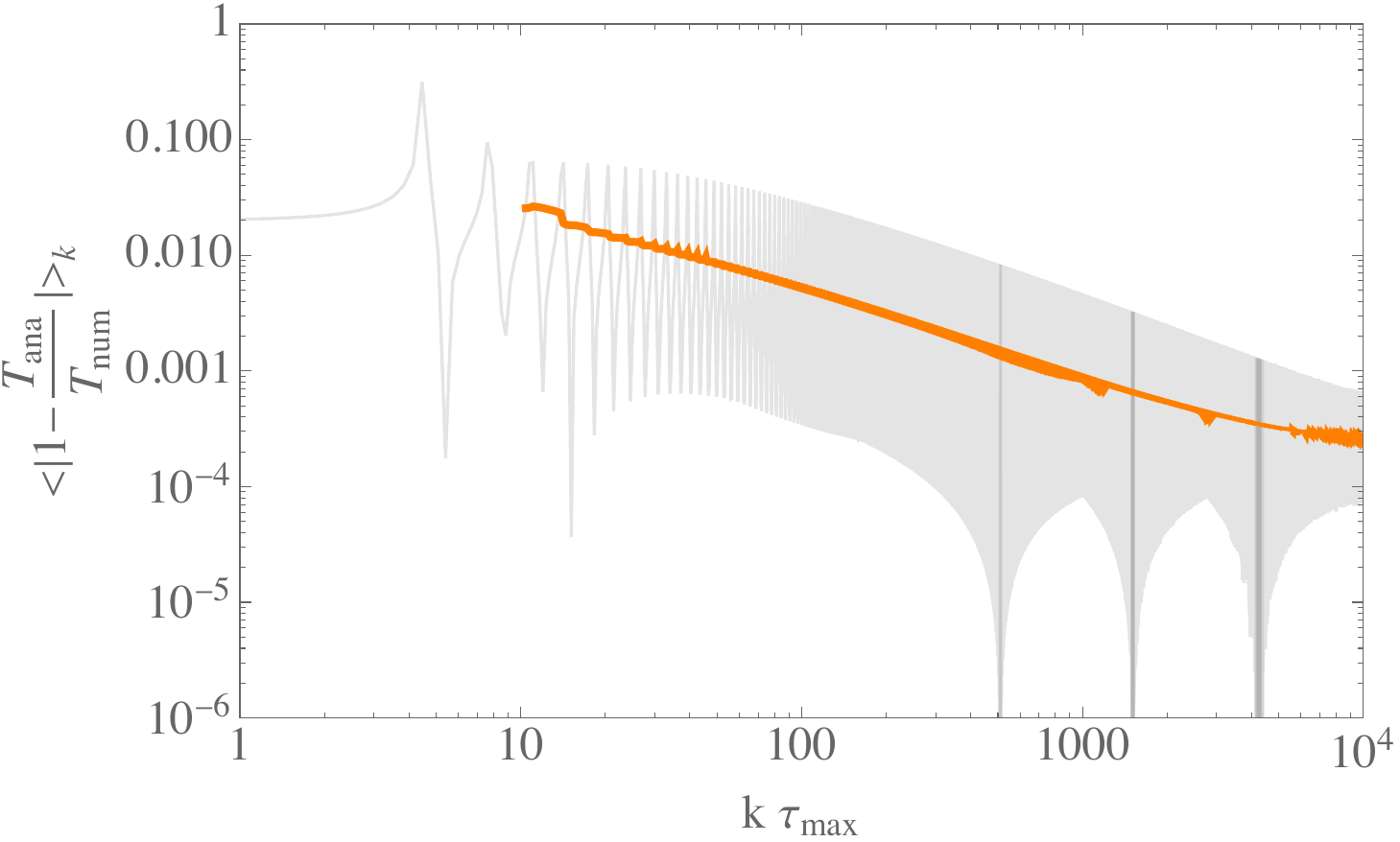}
       \caption{}
        \label{fig:tmat}
    \end{subfigure}
    \hfill
    \begin{subfigure}[b]{0.48\textwidth}
        \centering
       \includegraphics[height=4.4cm]{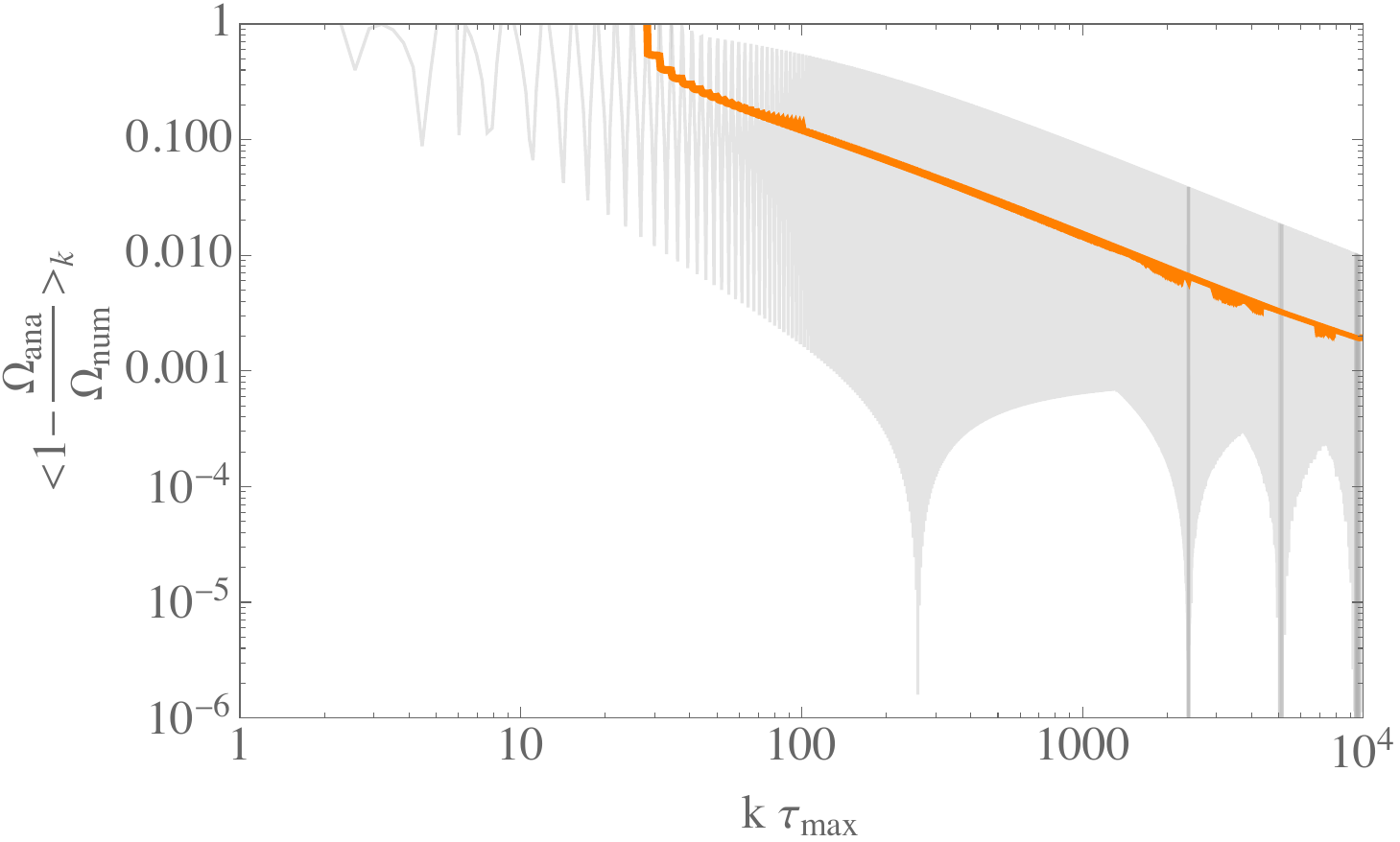}
       \caption{}
        \label{fig:ommat}
    \end{subfigure}
    \caption{Relative difference of the analytic and numeric transfer function and $\Omega_{\text{GW}}$ in a viscous matter epoch evaluated at $\tau=\tau_{\text{max}}$, for $\alpha=1$ and $\delta_{\text{max}}=10^{-2}$. In gray the relative difference, in orange the $k$ average.}
    \label{fig:mat}
\end{figure}
For all modes that re-entered the horizon during this era ($k\gtrsim\tau_{\text{max}}^{-1}$), the $k$ dependent damping of the transfer function is preserved during the following cosmic phases. Once the viscous phase ends, for $\tau>\tau_{\text{max}}$, the same argument presented in the previous section can be applied. Recalling that $j_1(k\tau)\sim-\cos(k\tau)/(k\tau)$, the transfer function for $\tau>\tau_\text{max}$, evolves as 
\begin{equation}
\mathcal{T}(k,\tau>\tau_{\text{max}})\simeq\mathcal{T}_{2}(k,\tau_{\text{max}})\frac{\cos(k\tau)}{\cos(k\tau_\text{max})}\frac{a(\tau_{\text{max}})}{a(\tau)}.
\label{eq:66}
\end{equation}
Hence, the fractional difference between the viscous and non-viscous transfer functions remains constant for $\tau>\tau_{\text{max}}$, taking its maximum value
\begin{equation}
\Delta_{\mathcal{T}}\simeq\frac{2\delta(\tau^{\text{exit}}(k))}{\alpha}\qquad,\qquad\tau>\tau_{\text{max}}\,.
\label{eq:67}
\end{equation}

The same argument applies to the normalized spectral energy density. Under the approximation $\mathcal{T}^{\prime}(k,\tau)\simeq k\,\mathcal{T}(k,\tau)$ valid for sub-horizon modes, we find
\begin{equation}
\Delta_\Omega\simeq
\frac{4\delta(\tau^{\text{exit}}(k))}{\alpha}\qquad,\qquad\tau>\tau_{\text{max}}\,.
\label{eq:68}
\end{equation}
This expression shows that, as in the radiation-dominated era, the impact of a small shear-viscous component during the matter-dominated phase translates into a permanent, in general $k$-dependent suppression of the transfer function and the corresponding spectral energy density by a relative factor proportional to $\delta(\tau^{\text{vis}}(k))/\alpha$. Also in this case the resulting effect on the spectral tilt is a slight increase on the tensor index. In the non-viscous case we have $n_0=n_t-2$, where $n_t\equiv\frac{d\ln\mathcal{P}}{d\ln k}$, while with the inclusion of viscosity and in the small friction term limit we find
\begin{equation}
n_{\text{eff}}\simeq n_0-4 \delta(\tau^{\text{exit}}(k))\frac{d\ln\tau^{\text{exit}}(k)}{d\ln k},
\end{equation}
where the additional term proportional to the friction is positive due to the decreasing behavior of $\tau^{\text{exit}}(k)$.

In Fig.~\ref{fig:mat} we show the relative difference between the  numerical and analytic approximation  for the transfer function given by Eq.~\eqref{eq:62} and the normalized energy density Eq.~\eqref{eq:64} at the end of the viscous era $\tau=\tau_{\text{max}}$, with a linear viscous scaling, and under the assumption $k/a \lambda_{\text{mfp}}^{-1}\to0$. As for the radiation case, this confirms that the analytical approximation is at the sub-percent level in the deep sub-horizon regime. These results can be directly extended to any viscous contribution arising in a dust-like fluid, such as cold dark matter or pre-recombination strongly coupled fluids i.e. dark fluids or more conventional electron-photon-baryon plasma, as we are going to show in the next section.

\section{Viscous damping in the electron-photon-baryon plasma}\label{sec7}
In this section we estimate the effect of the electron-photon-baryon fluid on the primordial gravitational wave (pGW) spectrum. For redshifts $z \gtrsim z_{\mathrm{rec}} \approx 10^{3}$, i.e.\ before recombination, photons and baryons are tightly coupled through Thomson scattering. The interaction rate is much larger than the Hubble expansion rate, and the mixture can be treated as a single relativistic fluid. In this regime the electron-photon-baryon plasma behaves effectively as a viscous fluid characterized by a non-zero shear viscosity $\eta$, which accounts for momentum diffusion due to the finite photon mean free path. Throughout this section we will assume the bulk viscosity to be negligible.

As explicitly shown in \cite{Pavone:2025vru}, shear viscosity does not modify the homogeneous FLRW background evolution, and therefore it affects cosmology only at the level of perturbations.  Thus, in the tightly coupled regime, the dominant impact of the electron-photon-baryon fluid on the pGW spectrum arises through shear-viscous damping.

\subsection{Shear viscous friction estimation}
The explicit expression for the shear viscosity can be obtained from Eq.~\eqref{visgen} and reads \cite{Dodelson:2003ft, Pajer:2012qep}
\begin{equation}
\eta=\frac{16}{45}\,\rho_{\gamma}\,t_{\gamma}
\qquad,\qquad 
t_{\gamma}=\frac{1}{\sigma_{T}n_{e}}=\lambda_{\text{mfp}}\,,
\label{eq:eta_def}
\end{equation}
where $\rho_\gamma$ is the photon energy density, $\sigma_{T}$ is the Thomson cross section and $n_{e}$ is the electron number density and $t_\gamma=\lambda_{\text{mfp}}$ is the relaxation time (mean-free-path) of interacting photons. Since $n_{e}\sim a^{-3}$ and $\rho_{\gamma}\sim a^{-4}$, then $\eta\sim a^{-1}$. From Eq.~\eqref{fricvis}, we obtain that the friction term is
\begin{equation}
\delta(\tau)=\lambda_{p}^{2}\eta\,a\,\mathcal{H}^{-1},
\label{eq:delta_def}
\end{equation}
so that $\delta(\tau)\sim \mathcal{H}^{-1}\sim \tau$ during both radiation domination and matter domination. We now proceed with the estimation of all the relevant physical quantities needed to evaluate the magnitude of the shear viscous parameter $\delta(\tau)$.

The photon energy density is
\begin{equation}
\rho_{\gamma}=\Omega_{\gamma}\rho_{cr}\,a^{-4}
\qquad,\qquad 
\rho_{cr}=\frac{3H_{0}^{2}}{\lambda_{p}^{2}}\,.
\label{eq:rho_defs}
\end{equation}
Using $H_{0}=100\,h\,\mathrm{km\,s^{-1}\,Mpc^{-1}}$ and $\Omega_{\gamma}h^{2}=2.47\times 10^{-5}$ \cite{Planck2018}, we find
\begin{equation}
\rho_{\gamma}\approx\frac{8.24\times 10^{-12}}{\lambda_{p}^{2}}\,a^{-4}\,\mathrm{Mpc}^{-2}.
\label{eq:rho_num}
\end{equation}
Moreover, the Thomson scattering rate per unit length is \cite{Dodelson:2003ft}
\begin{equation}
n_{e}\sigma_{T}=\sigma_T X_e \frac{\Omega_b\rho_{cr}}{m_p}\approx 5.2\times 10^{-7}\,a^{-3}\,\mathrm{Mpc}^{-1},
\label{eq:rate_num}
\end{equation}
where $X_{e}$ is the number of free electrons per hydrogen nucleus and it is $X_e\approx1$ up to recombination,  $\Omega_{b}h^{2}=0.0224$ and $m_p$ the proton mass. Inserting Eqs.~\eqref{eq:rho_defs} and \eqref{eq:rate_num} into \eqref{eq:eta_def} gives
\begin{equation}
\eta=\frac{16}{45}\,\frac{\Omega_\gamma\,m_p}{X_e\Omega_b\sigma_T}
\approx\frac{5.63\times 10^{-6}}{\lambda_{p}^{2}}\,a^{-1}\,\mathrm{Mpc}^{-1}.
\label{eq:eta_num}
\end{equation}
Therefore, using Eq.~\eqref{eq:delta_def},
\begin{equation}
\delta(\tau)=\lambda_{p}^{2}\eta\,a\,\mathcal{H}^{-1}
=\delta^*\,\mathcal{H}^{-1} ,
\label{eq:delta_tau}
\end{equation}
with $\delta^*\approx5.63\times 10^{-6}\,\text{Mpc}^{-1}$, i.e. the viscous parameter grows linearly with conformal time. This means that $\alpha=1$ in Eq.~\eqref{friction}. 

By writing the scale factor as follows
\begin{align}
a(\tau) &= a_{\rm eq}\,\frac{\tau}{\tau_{\rm eq}}
\quad\Rightarrow\quad
\mathcal{H}^{-1} = \tau\,,&\quad\text{for}\quad \tau \lesssim \tau_{\rm eq},\nonumber\\[6pt]
a(\tau) &= a_{\rm eq}\,\frac{(\tau+\tau_{\rm eq})^{2}}{4\,\tau_{\rm eq}^{2}}
\quad\Rightarrow\quad
\mathcal{H}^{-1} = \frac{\tau+\tau_{\rm eq}}{4}\,,&\quad\text{for}\quad \tau \gtrsim \tau_{\rm eq},\label{amat}
\end{align}
we find that the friction term can be written as
\begin{align}
\delta_{\text{rad}}(\tau)&=\delta^{\text{max}}_{\text{rad}}\left(\frac{\tau}{\tau_\text{eq}}\right)\,,&&\quad\text{for}\quad\tau<\tau_{\rm eq},\nonumber\\ 
\delta_{\text{mat}}(\tau)&=\frac{\delta^{\text{max}}_{\text{rad}}}{2}+\delta_{\text{mat}}^{\text{max}}\left(\frac{\tau}{\tau_\text{rec}}\right)\,,&&\quad\text{for}\quad\tau_{\text{eq}}\lesssim\tau\lesssim\tau_{\rm rec},\label{dmat}
\end{align}
where $\delta^{\text{max}}_{\text{rad}}\equiv\delta^*\tau_{\text{eq}}$, and $\delta^{\text{max}}_{\text{mat}}\equiv(\delta^*/2)\,\tau_{\text{rec}}$. 
The explicit values for the conformal time to recombination and equality follows from
\begin{equation}
\tau(a)=\int_{0}^{a}\frac{da}{a^{2}H(a)},
\qquad 
H(a)=H_{0}\sqrt{\Omega_{r}a^{-4}+\Omega_{m}a^{-3}},
\label{eq:tau_rec_int}
\end{equation}
so that
\begin{equation}
\tau=\frac{2}{H_{0}\Omega_{m}}
\Big[\sqrt{\Omega_{r}+\Omega_{m}a}-\sqrt{\Omega_{r}}\Big].
\label{eq:tau_rec_eval}
\end{equation}
Using $a_{\mathrm{eq}}=\Omega_{r}/\Omega_m$, with $\Omega_{m}=0.315$, $\Omega_{r}=9.2\times 10^{-5}$ and $H_{0}=67\,\mathrm{km\,s^{-1}\,Mpc^{-1}}$  \cite{Planck2018}, we have $\tau_{\text{eq}}\approx112\,\text{Mpc}$ and from  $a_{\mathrm{rec}}\approx10^{-3}$, $\tau_{\text{rec}}\approx300\,\mathrm{Mpc}$.
Inserting the numerical values into the definitions of the $\delta_{\text{rad}}^{\max}$ and $\delta_{\text{mat}}^{\max}$ we obtain

\begin{align}
\delta_{\text{rad}}^{\max}&\approx  6\times 10^{-4},\nonumber\\
\delta_{\text{mat}}^{\max}&\approx  1\times 10^{-3},\label{delta_max_mat}
\end{align}
confirming that viscous effects remain perturbatively small.

It is important to recall that viscosity is an effective description of the background fluid on physical length scales larger than the mean-free-path. From the arguments presented in previous section we have that  the modes able to probe the averaged fluid description of the photon fluid are those with $k\lesssim a\lambda_{\text{mfp}}^{-1}$, hence  $k\lesssim k_{\text{vis}}(\tau)$ with $k_\text{vis}(\tau)=k_{\text{mfp}}\,a^{-2}$, and $k_{\text{mfp}}\approx5.2\times10^{-7}\text{Mpc}^{-1}$. Using Eqs.~\eqref{amat}, we have
\begin{align}
    k_{\text{vis}}(\tau)&\simeq k_{\text{mfp}} \frac{\tau_{\text{eq}}^2}{a_{\text{eq}}^2\tau^2}\,,&&\quad\text{for}\quad\tau\lesssim\tau_{\text{eq}},\nonumber\\
    k_{\text{vis}}(\tau)&\simeq k_{\text{mfp}}\frac{16 \tau_{\text{eq}}^4}{a_{\text{eq}}^2(\tau+\tau_{\text{eq}})^4}\,,&&\quad\text{for}\quad\tau_{\text{eq}}\lesssim\tau\lesssim \tau_{\text{rec}}.\label{kvismat}
\end{align}

The peculiar structure of the above relations allow us to understand that for each mode we can define an exit time $\tau_{\text{vis}}^{\text{exit}}(k)$ from which the gravitational wave is not able to probe the the dissipative nature of the fluid. The only modes that feel the viscous phase for the whole duration are those with $k\lesssim k_{\text{vis}}(\tau_{\text{rec}})\approx0.54\,\text{Mpc}^{-1}$, hence for scales relevant to the CMB and LSS. Moreover it is useful for later convenience to compute the viscous freeze-out conformal time inverting the relation $k_{\text{mfp}}(\tau)=k$ as 
\begin{align}
\tau_{\text{vis}}^{\text{exit}}(k)&=\tau_{\text{vis}}^{\text{rad}}=\frac{\tau_\text{eq}}{a_{\text{eq}}}\sqrt{\frac{k_{\text{mfp}}}{k}}\,,&&\quad\text{for}\quad\tau\lesssim\tau_{\text{eq}}
\nonumber\\
\tau_{\text{vis}}^{\text{exit}}(k)&=\tau_{{\text{vis}}}^{\text{mat}}=-\tau_{\text{eq}}+2\tau_{\text{eq}}\left(\frac{k_{\text{mfp}}}{a_{\text{eq}}^2k}\right)^{1/4}\,,&&\quad\text{for}\quad\tau_{\text{eq}}\lesssim\tau\lesssim\tau_{\text{rec}},\label{exitmat}
\end{align}
where all the quantities are intended to be evaluated in $\text{Mpc}$, and moreover those expressions have to be intended valid only if $\tau_{\text{vis}}^{\text{rad}}\leq\tau_{\text{eq}}$ ($\tau_{\text{vis}}^{\text{mat}}\leq\tau_{\text{rec}}$) for the radiation (matter) epoch, otherwise the mode $k$ feels the viscous damping throughout its evolution. During these epochs, we model the viscosity as 
\begin{equation}
    \delta(\tau)=\delta^\star\mathcal{H}^{-1}\Theta(\tau_{\text{vis}}^{\text{exit}}(k)-\tau)\,.
\end{equation}
for those high frequency modes that ``exit'' the viscous phase before recombination.

\subsection{Impact on the pGW transfer function and normalized energy density power spectrum.}
For sub-horizon modes we distinguish modes that re-entered during the radiation epoch, hence $k\gtrsim \tau_{\text{eq}}^{-1}$, modes that re-entered during the matter dominated but still before recombination $\tau_{\text{rec}}^{-1}\lesssim k \lesssim \tau_{\text{eq}}^{-1}$ and modes that re-entered after recombination $k\lesssim \tau_{\text{rec}}^{-1}$. In addition we now have a second hierarchy of scales which are dictated by when a particular mode $k$ stops to feel the effective viscous description. We hence have the following:
\begin{itemize}
    \item $k\gtrsim k_{\text{vis}}(\tau_\text{eq})$, these modes re-enter during radiation domination and exit the viscous phase at $\tau_\text{vis}^{\text{rad}}$, hence before equality;
    \item $k_{\text{vis}}(\tau_\text{rec})\lesssim k\lesssim k_{\text{vis}}(\tau_\text{eq})$, these modes re-enter during radiation domination but exit the viscous phase between equality and recombination at $\tau_{\text{vis}}^{\text{mat}}$;
    \item $\tau_{\text{eq}}^{-1}\lesssim k \lesssim k_{\text{vis}}(\tau_\text{rec})$, these modes re-enter during radiation domination and are affected by the electron-photon-baryon plasma up to recombination;
    \item $\tau_{\text{rec}}^{-1}\lesssim k\lesssim \tau_{\text{eq}}^{-1}$, these modes re-enter during matter domination between equality and recombination;
    \item $k\lesssim \tau_{\text{rec}}^{-1}$, these modes re-enter after recombination.
\end{itemize}
\begin{figure}[ht!]
    \centering
    \includegraphics[width=15cm]{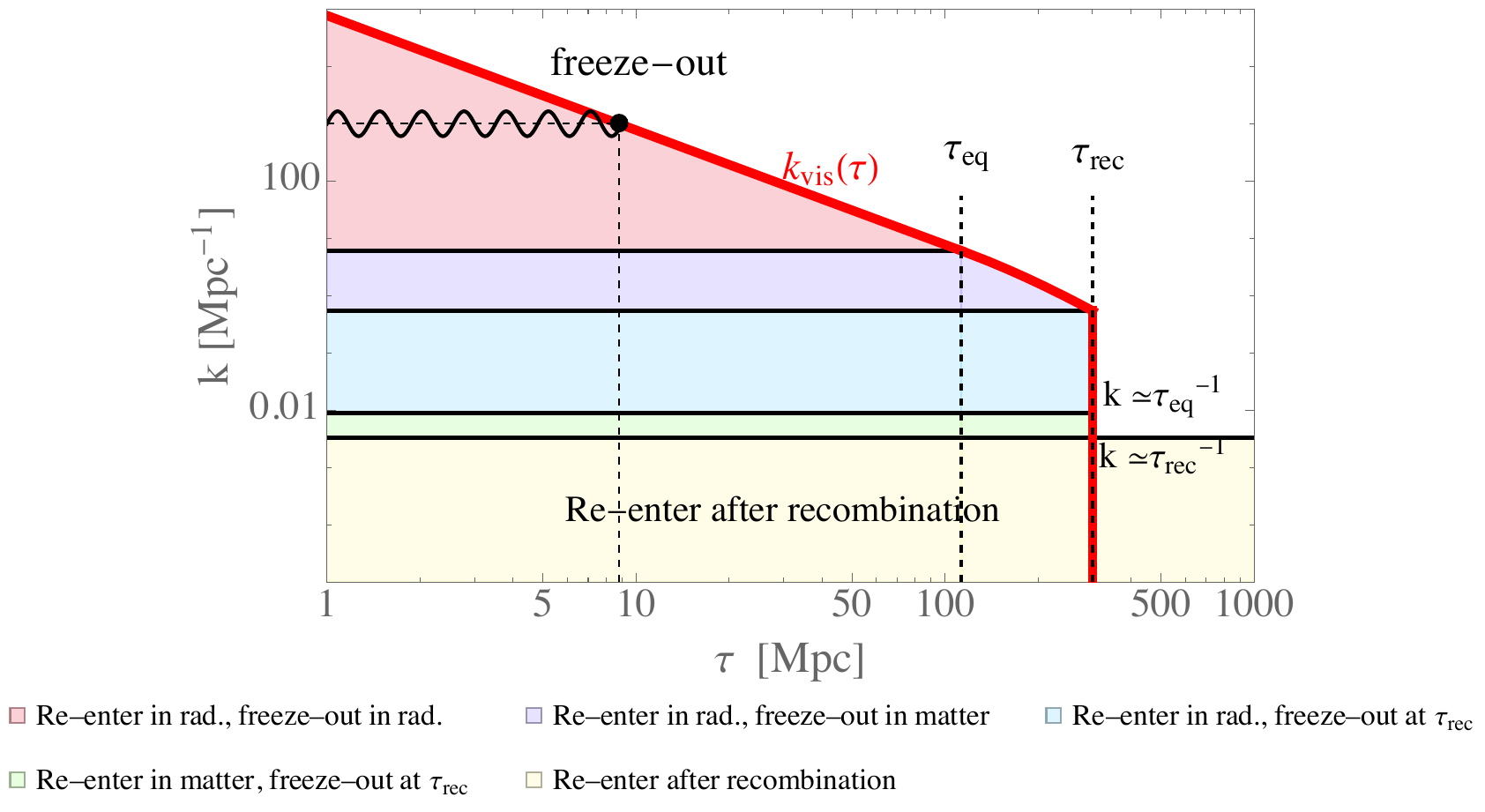}
    \caption{
    Evolution of the comoving viscous cutoff scale $k_{\rm vis}(\tau)$ (thick red curve).
    Modes below the curve ($k < k_{\rm vis}$) propagate within the hydrodynamic,
    shear-viscous regime of the photon-baryon-electron plasma and undergo
    dissipative damping, represented illustratively by the oscillatory wiggly mode.
    At the viscous freeze-out time $\tau_{\rm exit}(k)$ (black point), the mode exits the
    viscous regime and subsequently propagates in a non-viscous background. Horizontal and vertical markers correspond respectively to characteristic
    scales $k \simeq \tau_{\rm eq}^{-1}, \tau_{\rm rec}^{-1}$ and to matter-radiation  equality ($\tau_{\text{eq}}$) and recombination ($\tau_{\text{rec}}$). The colored  regions classify modes according to their re-entry time into the Hubble horizon and viscous freeze-out time.}
    \label{fig:explain}
\end{figure}

The structure above is visually summarized in Fig.~\ref{fig:explain}, which illustrates the hierarchy of re-entry and viscous freeze-out scales and the corresponding classification of modes.
The first modes to re-enter during  radiation domination $k\gtrsim k_{\text{vis}}(\tau_\text{eq})$, evolve on sub-horizon scales  and they exit the viscous phase before equality, in this case the transfer function takes the form (See Appendix \ref{appA})
\begin{equation}
\mathcal{T}(k,\tau_\text{eq})\simeq j_0(k\tau_\text{eq})\left(1-\frac{\delta_{\text{rad}}^{\text{max}}}{a_\text{eq}}\sqrt{\frac{k_{\text{mfp}}}{k}}\right),\qquad\qquad k\gtrsim k_{\text{vis}}(\tau_\text{eq}),
\end{equation}
where $k_{\text{vis}}(\tau_{\text{eq}})\approx 6.1\text{Mpc}^{-1}$. This result reinforce the statement already obtained in Eq.~\eqref{eq:59}, where  we concluded that the damping of the transfer function depends on the maximum value of the friction term \textit{per mode $k$}.\\
For these modes we can use Eq.~\eqref{eq:58} to conclude that for $k\gtrsim k_\text{vis}(\tau_\text{eq})$ and $\tau>\tau_{\text{eq}}$
\begin{equation}
\mathcal{T}(k,\tau)=\mathcal{T}(k,\tau_\text{eq})\frac{\sin(k \tau)}{\sin(k \tau_\text{eq})}\frac{a_\text{eq}}{a},
\end{equation}
and hence 
    \begin{align}
        \Delta_{\mathcal{T}} &\simeq\frac{\delta_{\text{rad}}^{\text{max}}}{a_\text{eq}}\sqrt{\frac{k_{\text{mfp}}}{k}},\label{visoutTrad} \\ 
        \Delta_\Omega &\simeq2\frac{\delta_{\text{rad}}^{\text{max}}}{a_\text{eq}}\sqrt{\frac{k_{\text{mfp}}}{k}}. \label{visoutOrad}     
    \end{align}
This result is in concordance with the result obtained in Eqs.~\eqref{eq:58} and~\eqref{eq:59a}, the spectrum slowly increases as $1-\sqrt{1/k}$ approaching the non-viscous result as $k\to\infty$, giving a $k$ dependent blue tilt to the spectrum. Assuming that the primordial gravitational-wave background follows a
power-law spectrum, $\Omega_{\rm GW}^{(0)}(k)\sim k^{n_0}$, we can
estimate how the viscosity affects its spectral index.
Introducing the correction
\begin{equation}
\Omega_{\rm GW}(k)=\Omega_{\rm GW}^{(0)}(k)
\left[1-A\,k^{-1/2}\right], \qquad
A\equiv 2\,\frac{\delta_{\rm rad}^{\max}}{a_{\rm eq}}\sqrt{k_{\text{mfp}}},
\end{equation}
the effective spectral index is given by
\begin{equation}
n_{\rm eff}(k)=
\frac{d\ln\Omega_{\rm GW}}{d\ln k}
=n_0+\frac{\tfrac{A}{2}k^{-1/2}}{1-Ak^{-1/2}}\,.
\end{equation}

For the limiting value $A\,k^{-1/2}\ll1$ which is satisfied when $k\to k_{\text{vis}}(\tau_{\text{eq}})$, we get that $A\,k^{-1/2}=2\delta_{\text{max}}^{\text{rad}}\ll1$. Hence, $n_{\rm eff}$ simplifies to
\begin{equation}
n_{\rm eff}(k)\simeq n_0+
\frac{\delta_{\rm rad}^{\max}}{a_{\rm eq}}
\sqrt{\frac{k_{\text{mfp}}}{k}}\,.
\end{equation}
Thanks to this result, we have that the damping term slightly increases the spectral tilt, making the spectrum mildly more blue on those scales, while the
effect rapidly decreases for larger $k$. The corresponding running in fact is
negative,
\begin{equation}
\alpha(k)\equiv\frac{dn_{\rm eff}}{d\ln k}
\simeq -\frac{1}{2}\,
\frac{\delta_{\rm rad}^{\max}}{a_{\rm eq}}
\sqrt{\frac{k_{\text{mfp}}}{k}},
\end{equation}
indicating a gradual return to the original slope at high frequencies. 

For modes with $k_{\text{vis}}(\tau_\text{rec})\lesssim k\lesssim k_{\text{vis}}(\tau_\text{eq})$, where $k_\text{vis}(\tau_\text{rec})\approx0.54 \,\text{Mpc}^{-1}$, we have at $\tau=\tau_{\text{eq}}$ that their transfer function can be written as (see Appendix \ref{appA})
\begin{equation}
    \mathcal{T}(k,\tau_{\text{eq}})\simeq j_0(k\tau_{\text{eq}})\left(1-\delta_\text{rad}^{\text{max}}\right).\label{radtilt}
\end{equation}
To evolve this solution up to $\tau>\tau_{\text{rec}}$, we introduce the modified pump-field $\tilde a$ and the Mukhanov-Sasaki-like variable as $v_k(\tau)\equiv\mathcal{T}(k,\tau)\tilde{a}$, which diagonalizes Eq.~\eqref{eq:15} and obtain the following  equation
\begin{equation}
    v''_k+\left(k^2-\frac{\tilde{a}''}{\tilde a}\right)v_k=0,\qquad\text{where}\qquad \frac{d}{d\tau}\ln\tilde a\equiv\frac{d}{d\tau}\ln a+\mathcal{H}\,\delta(\tau).
\end{equation}
Integrating the latter between $\tau_{\text{eq}}$ and $\tau\geq\tau_{\text{rec}}$, we find at the first order in $\delta^\star$
\begin{equation}
\frac{\tilde{a}}{\tilde{a}_{\text{eq}}}=\frac{a}{a_\text{eq}}\left[1+\delta^*\left(\tau_{\text{vis}}^{\text{mat}}-\tau_{\text{eq}}\right)\right],
\end{equation}
hence, following the same matching argument as in the viscous radiation case discussed in Sect.~\ref{sec5}, for $\tau\geq\tau_{\text{rec}}$ we find 
\begin{equation}
\mathcal{T}(k,\tau)=j_0(k,\tau_\text{eq})\frac{\sin(k\tau)}{\sin(k\tau_\text{eq})}\frac{a_\text{eq}}{a}\left[(1-\delta_\text{rad}^{\text{max}}\left(2\left(\frac{k_{\text{mfp}}}{a_{\text{eq}}^2k}\right)^{1/4}-1\right)\right],
\end{equation}
hence, when also $k_{\text{vis}}(\tau_{\text{rec}})\lesssim k\lesssim k_{\text{vis}}(\tau_{\text{eq}})$, we get 
\begin{equation}
        \Delta_{\mathcal{T}} \simeq\delta_\text{rad}^{\text{max}}\left[2\left(\frac{k_{\text{mfp}}}{a_{\text{eq}}^2k}\right)^{1/4}-1\right]\qquad\text{and}\qquad\Delta_\Omega\simeq2\,\Delta_\mathcal{T}\,.\label{visoutOmat}
\end{equation}

In the range $k_{\rm vis}(\tau_{\rm rec})\lesssim k \lesssim k_{\rm vis}(\tau_{\rm eq})$, the effective spectra acquires a blue-tilt and  for $\delta_{\text{rad}}^{\text{max}}\ll1$  
\begin{equation}
n_{\rm eff}(k)\simeq 
n_0+
\delta_{\rm rad}^{\max}
\left(\frac{k_{\text{mfp}}}{a_{\rm eq}^2 k}\right)^{1/4},
\end{equation}
which reaches its maximum at $k\simeq k_{\rm vis}(\tau_{\rm rec})$, with
\begin{equation}
n_{\rm eff}^{\max}\simeq 
n_0+
\delta_{\rm mat}^{\max}+\frac{\delta_{\text{rad}}^{\text{max}}}{2},
\end{equation}
and its minimum at $k_{\text{vis}}(\tau_\text{eq})$
\begin{equation}
    n_{\rm eff}^{\max}\simeq 
n_0+\delta_{\text{rad}}^{\max}.
\end{equation}
The running is negative and given by
\begin{equation}
\alpha(k)\equiv\frac{dn_{\rm eff}}{d\ln k}
\simeq
-\frac{1}{4}\,
\delta_{\rm rad}^{\max}
\left(\frac{k_{\text{mfp}}}{a_{\rm eq}^2 k}\right)^{1/4},
\end{equation}
showing that the spectrum gradually returns to the non-viscous slope at larger $k$.

For modes that re-enter during radiation domination but  are able to probe the viscous nature of the fluid up to recombination i.e. $\tau_{\text{eq}}^{-1}\lesssim k \lesssim k_{\text{vis}}(\tau_\text{rec})$, we have that for $\tau>\tau_{\text{rec}}$ similar arguments lead to the transfer function 
\begin{equation}
\mathcal{T}(k,\tau)\simeq j_0(k\tau_\text{eq})\left(1-2\delta_{\text{mat}}^{\text{max}}\right)\frac{a_{\text{eq}}}{a},
\end{equation}
hence
\begin{equation}
   \Delta_{\mathcal{T}} \simeq2\delta_{\text{mat}}^{\text{max}}\qquad\text{and}\qquad\Delta_\Omega \simeq2\,\Delta_\mathcal{T}\label{matdamp2}\\
\end{equation}
so, the only effect is a constant damping of the amplitude of the spectrum.

For modes that re-entered between matter-radiation equality and recombination i.e. $\tau_{\text{rec}}^{-1}\lesssim k\lesssim \tau_{\text{eq}}^{-1}$, we find that even taking into account the viscous super-horizon evolution during the radiation epoch, the result of Eq.~\eqref{eq:66} still applies (See Appendix \ref{appB}) and we conclude that the net result is the same constant damping as in Eqs.~\eqref{matdamp2}.

Finally, for modes that re-enter after $\tau_{\text{rec}}$ i.e. $k\ll\tau_{\text{rec}}^{-1}$, we have to take into account the super-horizon evolution during the viscous radiation and matter epochs. However the super-horizon evolution of the transfer function adds a correction of order $\delta (k\,\tau_{\text{rec}})^2$ (See Appendix \ref{appB}). Finally, for $k\tau_{\text{rec}}\ll1$, these evolve on super-horizon scales for the whole viscous phase, hence the correction to the transfer function is given by the super-horizon evolution, hence of order $\mathcal{T}(k,\tau_{\text{rec}})\sim1+\,O((1+\delta)(k\tau_{\text{rec}})^2)$, hence it is negligible.
   \begin{figure}[ht!]
    \centering
    \includegraphics[width=0.7\textwidth]{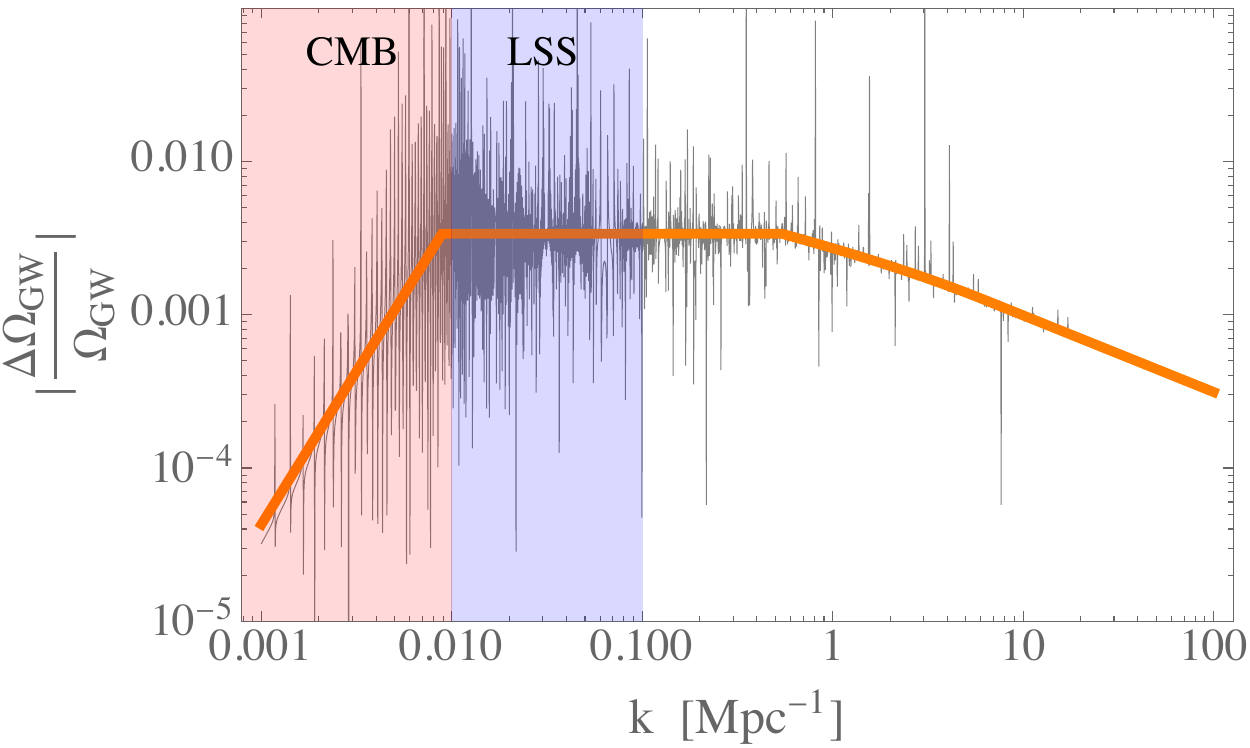}
    \caption{In light gray, the relative difference between the first order analytic approximation viscous $\Omega_{\text{GW}}$ and the analytic non-viscous case.  In orange the expressions Eqs.~\eqref{visoutOrad}, \eqref{visoutOmat} and \eqref{matdamp2}. The $\sim k^2$ for $k\to0$ is added to account for the viscous super-horizon suppression.}
    \label{fig:deltaOO}
\end{figure} 

As a final remark we highlight that the integral appearing in Eqs.~\eqref{eq:57} and \eqref{intmat}, characterizing the viscous correction to the transfer function in the radiation (matter) era, are solvable in closed form but not particularly enlightening. To this aim in Fig~\ref{fig:deltaOO}, we show the relative difference between the full analytic calculation for the viscous $\Omega_{\text{GW}}$, obtained by the explicit integration of the first order viscous correction, with the analytic non-viscous case. What can be concluded is that up to  phases, which introduce the spikes, the simple analytic results found here are robust.

\section{Conclusion}\label{conclusion}
In this work, we have analyzed the propagation of primordial gravitational waves after horizon re-entry in the presence of a shear-viscous cosmological background. Without assuming a specific inflationary scenario, we focused on the evolution of tensor modes during radiation- and matter-dominated epochs, when shear viscosity may arise as an effective description of imperfect fluids. We derived a general analytical framework to compute the transfer function and the corresponding normalized spectral energy density in the presence of a viscosity-induced friction term.

For a constant viscosity-to-Hubble ratio, $\delta = H_V/H$, we obtained exact solutions in terms of spherical Bessel functions and showed that viscosity modifies the standard $a^{-1}$ dilution law of tensor amplitudes in $a^{-(1+\delta)}$, leading to faster decay and an additional red tilt of the GW spectrum $\Omega_\text{GW}\sim k^{n_0+\Delta n}$, with $\Delta n=-2\beta\delta$. 

When $\delta(\tau)$ is time-dependent but small, we developed a controlled perturbative expansion and found that a transient viscous phase produces a frozen suppression of the amplitude of sub-horizon modes once the viscosity switches off. The magnitude of this suppression depends only on the  value of $\delta(\tau)$ experienced by each mode at the viscous freeze-out time $\tau_{\text{vis}}^{\text{exit}}(k)$ and remains imprinted during subsequent cosmological evolution. Moreover we found that this mechanism can generate a blue-tilt due to the different viscous freeze-out time per mode $k$.

We then applied this formalism to the physically relevant case of the tightly coupled photon-baryon-electron plasma before recombination. We computed the shear viscosity and the associated friction parameter $\delta(\tau)$ from first principles, showing that it grows linearly in conformal time. We identified a physical viscous freeze-out wavenumber, $k_{\text{vis}}(\tau)$, beyond which modes exit the hydrodynamical description because their wavelength becomes shorter than the photon mean-free-path.  Accounting for this scale-dependent exit time, we demonstrated that primordial GWs acquire a permanent damping at the level of $\sim 10^{-3}$ in $\Omega_{\text{GW}}$ with features characterized  by a running tensor index $n_{\text{eff}}$ for CMB-LSS relevant modes, while the impact rapidly decreases for larger wavenumbers.

Our results confirm the robustness of the standard prediction for primordial gravitational waves in $\Lambda$CDM: the shear viscosity of the electron-photon-baryon plasma does not significantly alter the observable spectrum and the modification is too small and at too low frequencies to have relevant consequences for future gravitational antennas detection capability of a pGW background. However, the analytical framework developed here provides a general tool to describe non-standard early-Universe scenarios involving dissipative epochs, such as dark radiation with self-interactions, warm inflation, reheating dynamics, or strongly coupled hidden sectors. In these contexts, shear-viscous damping may leave testable signatures in the spectral shape and amplitude of the pGW background, potentially accessible to pulsar-timing arrays and future space-based detectors such as LISA, DECIGO and ET. On a complementary side, the modification to the transfer function could alter the spectrum of late-time observables, such as the lensing shear and magnification as sourced by tensor perturbations \cite{Fanizza:2022wob}. The impact on those could provide an independent probe, possibly accessible in the next future.

\section*{Acknowledgements}
EP and LT are supported in part by INFN under the program TAsP: \textit{``Theoretical Astroparticle Physic''}. EP and LT are also supported by the research grant number 2022E2J4RK ``PANTHEON: Perspectives in Astroparticle and Neutrino THEory with Old and New messenger'', under the program PRIN 2022 funded by the Italian Ministero dell'Universit\`a e della Ricerca (MUR) and by the European UnionNext Generation EU. GF acknowledges the COST Action CosmoVerse, CA21136, supported by COST (European Cooperation in Science and Technology). GF is also member of the Gruppo Nazionale per
la Fisica Matematica (GNFM) of the Istituto Nazionale di Alta Matematica (INdAM).

\appendix
\section{Analytical approximation for linear $\delta$ for modes with $k\lesssim \tau_{\text{eq}}^{-1}$} \label{appA}
In this appendix we provide the analytic solution for the transfer function for a linear $\delta$ and the wavelength cutoff for modes that exit the viscous phase before recombination.

During radiation domination, i.e. $\tau\lesssim\tau_{\text{eq}}$, we have from equation Eqs.~\eqref{eq:57} and first of Eqs~\eqref{dmat}
\begin{equation}
\mathcal{T}_{1}^{(1)}(k,\tau)=-\frac{2}{(k\tau_{\text{eq}})}\delta^{\text{max}}_\text{rad}\int_{0}^{k\tau^\star_{\text{rad}}}d\xi\,\xi^{2}\bigg[j_{0}\left(k\tau\right)y_{0}\left(\xi\right)-y_{0}\left(k\tau\right)j_{0}\left(\xi\right)\bigg]j_{1}\left(\xi\right),\qquad\quad\tau^{\star}_{\text{rad}}=\text{min}(\tau,\tau_\text{vis}^{\text{rad}}) \label{visradcut}
\end{equation}
where $\tau_\text{vis}^{\text{rad}}$ is given by first of Eqs.~\eqref{exitmat}. The previous expression can be analytically integrated to give
\begin{equation}
\begin{aligned}
\mathcal{T}_{1}(k,\tau)&=j_0(k,\tau)-\frac{\delta_{\mathrm{rad}}(\tau)}{2 (k\tau)^2}
\Big[
\cos(k\tau)\!\big(-2\,\text{Ci}(2k\tau^\star_{\text{rad}}) + 2\log(k\tau^\star_{\text{rad}}) + 2\gamma - 1 + \log 4 \big)\\
&+ \sin(k\tau)\!\left[-2\,\text{Si}(2k\tau^\star_{\text{rad}}) + 2k\tau^\star_{\text{rad}}\right]
+ \cos\!\big(k(\tau - 2\tau^\star_{\text{rad}})\big)
\Big]. \label{radexact}
\end{aligned}
\end{equation}
where the trigonometric integral functions are defined as:
\begin{equation}
\text{Si}(x) = \int_0^x \frac{\sin t}{t}\, dt,
\qquad
\text{Ci}(x) = -\int_x^{\infty} \frac{\cos t}{t}\, dt,
\end{equation}
and
\begin{equation}
\gamma = \lim_{n \to \infty} \left( \sum_{k=1}^n \frac{1}{k} - \log n \right)
\simeq 0.57721
\end{equation}
is the Euler-Mascheroni constant. For sub-horizon modes ($k\tau_{\text{eq}}\gg1$) at the leading order in $1/k\tau$ we find accordingly with Eq.~\eqref{eq:57a}
\begin{equation}
    \mathcal{T}_1(k,\tau)\simeq j_0(k\tau)\left(1-\delta_\text{rad}(\tau^\star_{\text{rad}})\right).\label{radtilt}
\end{equation}
It is pivotal to stress that for $\tau_{\text{vis}}^{\text{rad}}<\tau$, then $\tau^\star_{\text{rad}}=\tau_{\text{vis}}^{\text{rad}}$ and so it depends on $k$ via Eqs.~\eqref{exitmat}. In particular at $\tau_{\text{eq}}$, we have that the transfer function acquires the following form 
\begin{equation}
\begin{aligned}
\mathcal{T}_1(k,\tau_\text{eq})&\simeq j_0(k\tau_\text{eq})(1-\delta_\text{rad}^{\text{max}})\qquad\qquad &k\lesssim &k_\text{vis}(\tau_\text{eq})\\ \\
\mathcal{T}_1(k,\tau_\text{eq})&\simeq j_0(k\tau_\text{eq})\left(1-\frac{\delta_{\text{rad}}^{\text{max}}}{a_\text{eq}}\sqrt{\frac{k_{\text{mfp}}}{k}}\right)&k\gtrsim&k_{\text{vis}}(\tau_\text{eq})
\end{aligned}
\end{equation}

\section{Analytical approximation for linear $\delta$ for modes with $k\gtrsim{\tau_{\text{eq}}}^{-1}$ } \label{appB}

The modes $k\lesssim \tau_{\text{rec}}^{-1}$,  re-enter the horizon during the matter dominated epoch and they are affected by the viscosity up to $\tau_{\text{rec}}$, during they super-horizon evolution. Using Eq.~\eqref{dmat}, the evolution equation for the transfer function in terms of $\tilde\tau=\tau+\tau_{\text{eq}}$ can be written as
\begin{align}
    \mathcal{T}_2''(k,\tilde\tau)+\frac{4}{\tilde{\tau}}\Big( 1+\delta_{\text{mat}}^{\text{max}}\frac{\tilde\tau}{\tau_{\text{rec}}}\Theta(\tau_{\text{rec}}&+\tau_\text{eq}-\tilde\tau)\Theta(\tau_{\text{vis}}^{\text{mat}}+\tau_{\text{eq}}-\tilde\tau) \Big)\mathcal{T}_2^\prime(k,\tilde{\tau})+k^2\mathcal{T}_2(k,\tilde{\tau})=0,\\
    \nonumber\\
\text{with}\qquad\mathcal{T}_2(k,2\tau_\text{eq})&=\mathcal{T}_1(k,\tau_\text{eq})
\qquad\text{and}\qquad
\mathcal{T}_2'(k,2\tau_\text{eq})=\mathcal{T}_1^\prime(k,\tau_{\text{eq}})\,,\nonumber
    \end{align}
where the matching conditions have been chosen to match the solution at the end of radiation domination given by Eq.~\eqref{radexact} and with $\tau^\star_\text{rad}=\tau$. Using Eqs.~\eqref{eq:34-1}, \eqref{eq:30} into Eq.~\eqref{eq:40} and upon the substitution $\xi=k\tau'$, we have for $\tau_{\text{eq}}\lesssim\tau\lesssim\tau_{\text{rec}}$, that the first order correction is
\begin{align}
    \mathcal{T}_2^{(1)}(k,\tilde\tau)&=-\frac{4\delta_{\text{mat}}^{\text{max}}}{(k\,\tau_{\text{rec}})(k\,\tilde{\tau})}\int_{2\xi_{\text{eq}}}^{\xi^\star_\text{mat}}d\xi\,\xi^2[j_1(k \tilde{\tau})y_1(\xi)-y_1(k\tilde\tau)j_1(\xi)]\times[C_1\,j_2(\xi)+C_2\,y_2(\xi)],\label{intmat}\\\nonumber \\ 
    &\text{where}\qquad\xi_\text{eq}\equiv k\tau_\text{eq}\quad,\quad \xi_\text{max}^\star\equiv k\tau_\text{mat}^\star\quad,\quad\tau_\text{mat}^\star\equiv\text{min}(\tau_{\text{eq}}+\tau,\tilde\tau_{\text{vis}}^\text{mat})\,,\nonumber\\
&\qquad\qquad\tilde\tau_{\text{vis}}^\text{mat}=\tau_\text{eq}+\tau_{\text{vis}}^{\text{mat}}=2\tau_\text{eq}\left(\frac{k_{\text{mfp}}}{a^2_\text{eq}k}\right)^{1/4}\,,\nonumber
    \end{align}
and $C_1$ and $C_2$ are obtained by the leading order matching
\begin{equation}
    \mathcal{T}_2^{(0)}(k,2\tau_\text{eq})=\mathcal{T}_1(k,\tau_\text{eq}),\qquad\qquad\mathcal{T}_2^{(0)\prime}(k,2\tau_\text{eq})=\mathcal{T}^\prime_1(k,\tau_\text{eq}),
\end{equation} 
where $\mathcal{T}^{(0)}_2(k,\tau)$ is given by Eq.~\eqref{eq:32-1}, and $\mathcal{T}_1(k,\tau)$ by Eq.~\eqref{radexact}.

Even though it is possible to evaluate in closed form both the integral and the coefficients $C_i$, the expression is lengthy and gives no particular insight, even though it accounts for the exact phase shift in the transfer function. Being interested in the overall amplitude we want to find a simple approximated form which takes into account the super-horizon damping during the radiation evolution. Upon expanding Eq.~\eqref{radexact} for $k\tau\ll1$ , we obtain
\begin{equation}
  \mathcal{T}_1(k,\tau)=1-\frac{(k\tau)^2}{6}\left(1-\frac{\delta_{\text{rad}}^{\text{max}}}{3}\frac{\tau}{\tau_{\text{eq}}}\right)+O((k\tau)^3). \label{sup-hor}
\end{equation}
We assume that our modes are still super-horizon at equality, hence $k \tau_{\text{eq}}\ll1$ and evaluating the matching coefficients in this regime we find
\begin{equation}
\begin{aligned}
    C_1&=3+\left(\frac{5}{6}+\frac{\delta_{\text{rad}}^{\text{max}}}{2}\right)(k\tau_{\text{eq}})^2+O((k\tau_{\text{eq})}^4),\\
    C_2&=O((k\tau_{\text{eq}})^5,
    \end{aligned}
\end{equation}
so using the same averaging over oscillation procedure shown in Sect.\ref{sec5} (averaging over oscillations, keeping the leading order in $\xi$ in the integral and in the $k\tau_{\text{eq}}\ll1$ limit and $k\tau_{\text{rec}}\gg1$), we find that Eq.~\eqref{intmat}  can be approximated as 
\begin{equation}
\mathcal{T}_2^{(1)}(k,\tilde\tau)\simeq-\frac{2\delta_{\text{mat}}^{\text{max}}}{(k\,\tau_\text{rec})}C_1\,{j_1(k\tilde{\tau} )},
\end{equation}
hence for $\tau_{\text{eq}}\lesssim\tau\lesssim\tau_{\text{rec}}$ and $\tau_{\text{rec}}^{-1}\lesssim k\lesssim \tau_{\text{eq}}^{-1}$ the transfer function reads
\begin{equation}
\mathcal{T}_2(k,\tau)\simeq C_1 \frac{j_1(k\tau)}{k\tau}\left(1-2\delta^{\text{max}}_{\text{mat}}\frac{\tau}{\tau_\text{rec}}\right).
\end{equation}
We remark that this approximated expression has been explicitly checked solving the integral and matching conditions and performing the limits $k\tau_{\text{eq}}\ll1$ and $k\tau\gg1$. Using the same arguments has in Sect.\ref{sec6}, we conclude that the fractional difference is given by Eqs.~\eqref{matdamp2}.

Final for $k\tau_{\text{rec}}\ll1$, these evolve on super-horizon scales for the whole viscous phase, hence the correction to the transfer function is given by the super-horizon evolution Eq.~\eqref{sup-hor}, hence of order $\delta_{\text{rad}}^{\max} (k\,\tau_{\text{rec}})^2$ (or $\delta_{\text{mat}}^{\max} (k\,\tau_{\text{rec}})^2$) in radiation (or matter), hence it is negligible.

\bibliography{biblio}
\bibliographystyle{JHEP}

\end{document}